\documentstyle[prd,aps,twocolumn]{revtex}
\begin{document}
\draft

\twocolumn[\hsize\textwidth\columnwidth\hsize\csname @twocolumnfalse\endcsname
\title{Toward Quantum Gravity II: Quantum Tests}
\author{Heui-Seol Roh\thanks{e-mail: hroh@nature.skku.ac.kr}}
\address{BK21 Physics Research Division, Department of Physics, Sung Kyun Kwan University, Suwon 440-746, Republic of Korea}
\date{\today}
\maketitle

\begin{abstract}
This study toward quantum gravity (QG) introduces an $SU(N)$ gauge theory with the
$\Theta$ vacuum term for gravitational interactions, which leads to a group $SU(2)_L
\times U(1)_Y \times SU(3)_C$ for weak and strong interactions through dynamical
spontaneous symmetry breaking (DSSB). Newton gravitation constant $G_N$ and the
effective cosmological constant are realized as the effective coupling constant and
the effective vacuum energy, respectively, due to massive gauge bosons. A gauge theory
relevant for the non-zero gauge bosons, $\simeq 10^{-12}$ GeV, and the massless gauge
boson (photon) is predicted as a new dynamics for the universe expansion: this is
supported by the repulsive force, indicated in BUMERANG-98 and MAXIMA-1 experiments,
and cosmic microwave background radiation. Under the constraint of the flat universe,
$\Omega = 1 - 10^{-61}$, the large cosmological constant in the early universe becomes
the source of the exponential expansion in $10^{30}$ order as expected in the
inflation theory, nearly massless gauge bosons are regarded as strongly interacting
mediators of dark matter, and the baryon asymmetry is related to the DSSB mechanism.
\end{abstract}

 ] \narrowtext

\section{Introduction}

General relativity \cite{Eins} as presently accepted, classical theory or standard big
bang theory based on general relativity has outstanding problems: the singularity,
cosmological constant or vacuum energy, flat universe, baryon asymmetry, horizon
problem, the large scale homogeneity and isotropy of the universe, dark matter, galaxy
formation, discrepancy between astrophysical age and Hubble age, etc. According to
recent experiments, BUMERANG-98 and MAXIMA-1 \cite{Jaff}, the universe is flat and
there exists repulsive force represented by non-zero vacuum energy, which plays
dominant role in the universe expansion. The aim of this paper is to introduce quantum
gauge theory for gravitational interactions, which may resolve the problems in general
relativity toward the eventual unification of fundamental forces and may satisfy the
experiment results. In this paper, the quantum features of quantum gravity (QG) as a
gauge theory associated with a group $G$ are suggested even though the group $G$ is
exactly not known at present: the group chain is given by $G \supset SU(2)_L \times
U(1)_Y \times SU(3)_C$ where $G$, $SU(2)_L \times U(1)_Y$, and $SU(3)_C$ groups are
for gravitation, weak \cite{Glas}, and strong \cite{Frit} interactions respectively.
An $SU(N)$ gauge theory as a trial theory toward QG is specifically introduced to
resolve topics relevant for Newton gravitation constant $G_N$ and the cosmological
constant in the previous paper \cite{Roh1}. This paper then tries to concentrate on
quantum tests predicted by this scheme from the Planck scale $10^{-33}$ cm to the
universe scale $10^{28}$ cm.

There is no distinct connection between Einstein's general relativity on which the
standard cosmology is based and gauge theory on which grand unified theory (GUT) is
based. The incompatibility of the two modern theories, general relativity and gauge
theory, is thus the biggest obstacle to the unification of the two theories into one
theoretical framework; as known widely, the unification of the two theories has been
one of the greatest challenge in physics. There are usually two directions toward
quantum gravitation theory or the unification of fundamental forces: superstring
theory and Kaluza-Klein theory in the higher dimensions and the Planck scale.
Superstring theory is considered to be one of the most promising candidate in the
unification of forces but there has been no known compactification method to break
down to the real, low energy world and no clear answer to how superstring theory
solves the cosmological constant problem. In this context, it is quite natural to
develop quantum gauge theory to overcome these problems as well as to satisfy the
recent experiments \cite{Jaff}. As a step toward the super-grand unification of
fundamental forces or toward the systematic description of the universe evolution, the
$SU(N)$ gauge theory with the $\Theta$ vacuum term for gravitation is tested from
several viewpoints since compelling theoretical and experimental arguments for QG are
very appealing. The difficulty in renormalization or quantization for gravity may
disappear in this scheme since, whatever the group $G$ is, the quantization method of
gauge theory can be used to quantize gravitational wave and all gauge theories are
renormalizable \cite{Hoof}. The dimensionless coupling constant for a renormalizable
gauge theory is produced by interpreting that Newton gravitation constant acquires the
dimension of inverse energy square due to the graviton mass \cite{Roh1}. This work
suggests that the effective cosmological constant is related to the effective vacuum
energy represented by massive gauge bosons and then that the condensation of singlet
gravitons cancels the vacuum energy in a real world \cite{Roh1}.  As one of quantum
tests, a gauge theory for a new type of force with the gauge boson of the extremely
small mass $M_G \approx 10^{-12}$ GeV, which is associated with the observed
cosmological constant $\Lambda_0 = 8 \pi G_N M_G^4 \simeq 10^{-84} \ \textup{GeV}^2$,
is proposed in order to explain the expansion of the present universe; cosmic
microwave background radiation at $2.7$ K is a conclusive clue of massless gauge
bosons during DSSB. Other quantum tests such as inflation, candidates of dark matter,
baryon and lepton asymmetries, cosmological parameters, mass generation mechanism,
$\Theta$ constant and quantum numbers, conservation laws etc. are also suggested to
illustrate this scheme. The present work is mainly restricted to the low, real
dimensions of spacetime without considering supersymmetry.

This paper is organized as follows. In Section II, an $SU(N)$ gauge theory toward QG
is suggested without considering supersymetry and then DSSB is briefly introduced as
discussed in the previous paper \cite{Roh1}. In Section III, quantum tests toward QG
are suggested beyond Einstein's general relativity and the standard model. The
expansion of the universe is taken into account by a gauge theory with the nearly
massless gauge boson and the massless gauge boson, which is hinted by cosmic microwave
background radiation. Several other significant tests such as inflation, candidates of
dark matter, baryon and lepton asymmetries, cosmological parameters, mass generation
mechanism, $\Theta$ constant and quantum numbers, conservation laws, etc. are also
discussed. Section IV is devoted to conclusions.

\section{Toward Quantum Gravity}

An $SU(N)$ gauge invariant Lagrangian density with the $\Theta$ vacuum term is,
without taking into account supersymmetry, introduced
toward QG as a trial theory even though the exact group $G$ for gravity is not unveiled \cite{Roh1}.
DSSB triggered by the $\Theta$ term is adopted to generate gauge boson mass and fermion mass.
Natural units with $\hbar = c = k_B = 1$ are preferred for convenience throughout this paper unless otherwise specified.

The gauge invariant Lagrangian density is, in four vector notation, given by
\begin{equation}
\label{qchr}
{\cal L} = - \frac{1}{2} Tr  G_{\mu \nu} G^{\mu \nu}
+ \sum_{i=1}  \bar \psi_i i \gamma^\mu D_\mu \psi_i
\end{equation}
where the subscript $i$ stands for the classes of pointlike spinors, $\psi$ for the
spinor, and $D_\mu = \partial_\mu - i g_g A_\mu$ for the covariant derivative with the
gravitational coupling constant $g_g$. Particles carry the local charges and the gauge
fields are denoted by $A_{\mu} = \sum_{a=0} A^a_{\mu} \lambda^a /2$ with matrices
$\lambda^a$, $a = 0,.., (N^2-1)$. The field strength tensor is given by $G_{\mu \nu} =
\partial_\mu A_\nu - \partial_\nu A_\mu - i g_g [A_\mu, A_\nu]$. A current anomaly
\cite{Adle} is taken into account to show DSSB in analogy with the axial current
anomaly, which is linked to the $\Theta$ vacuum in QCD as a gauge theory
\cite{Hoof2,Roh3}. The bare $\Theta$ term is added as a single, additional
nonperturbative term to the Lagrangian density (\ref{qchr})
\begin{equation}
\label{thet}
{\cal L}_{QG} = {\cal L}_{P} + \Theta \frac{g_g^2}{16 \pi^2} Tr G^{\mu \nu} \tilde G_{\mu \nu},
\end{equation}
where ${\cal L}_{P}$ is the perturbative Lagrangian density (\ref{qchr}), $G^{\mu \nu}$
is the field strength tensor, and $\tilde G_{\mu \nu}$ is the dual of the field strength tensor.
Since the $G \tilde G$ term is a total derivative, it does not affect the perturbative aspects of the theory.

DSSB consists of two simultaneous mechanisms; the first mechanism is the explicit
symmetry breaking of gauge symmetry, which is represented by the gravitational factor
$g_f$ and the gravitational coupling constant $g_g$, and the second mechanism is the
spontaneous symmetry breaking of gauge fields, which is represented by the
condensation of gravitational singlet gauge fields.  Gauge fields are generally
decomposed by charge nonsinglet-singlet on the one hand and by even-odd discrete
symmetries on the other hand: they have dual properties in charge and discrete
symmetries. Four singlet gauge boson interactions in (\ref{thet}), apart from
nonsinglet gauge bosons, are parameterized by the $SU(N)$ symmetric scalar potential:
\begin{equation}
\label{higs}
V_e (\phi) = V_0 + \mu^2 \phi^2 + \lambda \phi^4
\end{equation}
which is the typical potential with $\mu^2 < 0$ and $\lambda > 0$
for spontaneous symmetry breaking. The first term of the right
hand side corresponds to the vacuum energy density representing
the zero-point energy by even parity singlets. The odd-parity
vacuum field $\phi$ is shifted by an invariant quantity $\langle
\phi \rangle$, which satisfies
\begin{equation}
\label{higs1}
\langle \phi \rangle^2 = \phi_0^2 + \phi_1^2 +  \cdot \cdot \cdot + \phi_{N}^2
\end{equation}
with the condensation of odd-parity singlet gauge bosons: $\langle \phi \rangle = (\frac{- \mu^2}{2
\lambda})^{1/2}$. DSSB is relevant for the surface term $\Theta \frac{g_g^2}{16 \pi^2}
Tr G^{\mu \nu} \tilde G_{\mu \nu}$, which explicitly breaks down the $SU(N)$ gauge
symmetry for quantum gravity through the condensation of odd-parity singlet gauge
bosons. The $\Theta$ can be assigned by an dynamic parameter by
\begin{equation}
\label{thev}
\Theta = 10^{-61} \ \rho_G /\rho_m
\end{equation}
with the matter energy density $\rho_m$ and the vacuum energy density $\rho_G = M_G^4$.
The detail of the $\Theta$ constant will be discussed in Section III.

The Newton gravitation constant as the effective gravitational coupling becomes
\begin{equation}
\label{glprr}
\frac{G_N}{\sqrt{2}} = - \frac{g_f g_g^2}{8 (k^2 - M_G^2 )} \simeq
\frac{g_f g_g^2}{8 M_G^2} \simeq 10^{-38} \ \textup{GeV}^{-2}
\end{equation}
and graviton as a gauge boson for gravitation has the Planck mass at the
Planck scale:
\begin{equation}
\label{grmsr}
M_G \approx M_{Pl} \approx 10^{19} \ \textup{GeV}
\end{equation}
which is reduced to a smaller value due to the condensation of the singlet graviton.
The gravitational factor $g_f$ is defined by $g_f = \frac{1}{4} (g_3^\dagger \lambda^a
g_1) (g_2^\dagger \lambda_a g_4)$ with gravitational charge fields, $g_i$ with $i = 1
\sim 4$, in analogy with the color factor $c_f$ in QCD. Note that the conventional
relation $G_N = 1 /M_{Pl}^2$ is adjusted to $G_N \simeq \sqrt{2} g_f g_g^2 /8
M_{Pl}^2$ in (\ref{glprr}). The gauge boson mass below the Planck energy can be cast
by
\begin{equation}
\label{gama} M_G^2 = M_{Pl}^2 - g_f g_g^2 \langle \phi \rangle^2 = g_f g_g^2
[A_{0}^2 - \langle \phi \rangle^2]
\end{equation}
with the even parity singlet gauge boson $A_{0}$, the odd-parity singlet gauge boson
condensation $\langle \phi \rangle$, the gravitational charge factor $g_f$, and the coupling
constant $g_g$. The gravitational factor $g_f$ used in (\ref{gama}) is the symmetric
factor for a gauge boson with even parity and the asymmetric factor for a gauge boson
with odd parity. This process makes the breaking of discrete symmetries P, C, T, and
CP.

The form of the Lagrangian density (\ref{thet}) may be analogously
used at lower energies as well as the Planck energy; the GWS model \cite{Glas}
as an $SU(2)_L \times U(1)_Y$ gauge theory at the weak scale and quantum chromodynamics (QCD)
as an $SU(3)_C$ gauge theory \cite{Frit} at the strong scale.
The number $N$ of an $SU(N)$ gauge theory will be related to the intrinsic
quantum number of the intrinsic space.
Based on the gauge theory (\ref{thet}) and dynamical
spontaneous symmetry breaking, quantum tests are discussed in the
following section.

\section{Quantum Tests}

In the previous section and paper \cite{Roh1}, Newton gravitation constant is defined
as the effective coupling constant and the effective cosmological
constant is connected to the effective vacuum energy due to
massive gauge bosons. In this section, a gauge theory responsible
for the universe expansion at the present epoch is proposed as a
new force and quantum tests beyond Einstein's general relativity
or the standard big bang theory in cosmology are discussed from
the view points of gauge theories from the Planck epoch to the
present epoch. Massive gauge bosons play central roles at all
times of the universe evolution. The newly introduced concepts are
the vacuum quantization with the maximum wavevector mode $N_R =
i/(\Omega -1)^{1/2} \approx 10^{30}$ or the total particle number
$N_G = 4 \pi N_R^3/3 \approx 10^{91}$, the constant $\Theta =
10^{-61} \ \rho_G/\rho_m$ defined from the flat universe
$\Omega - 1 = - 10^{-61}$, and the time scale $t = 1/H_e = (3/8 \pi G_N
M_G^4)^{1/2}$ expressed by the gauge boson mass $M_G$. Since QG
has the dimensionless coupling constant $\alpha_g$ and effective
coupling constant $G_N$ is acquired by DSSB, it is renormalizable
as asserted by 't Hooft \cite{Hoof}: the higher order terms of
$G_N$ do not diverge since the particle energy satisfies $E <
M_G$.

The constraint of the flat universe,
\begin{equation}
\Omega - 1 = 10^{-61} ,
\end{equation}
is required by quantum gauge theory and inflation scenario
and is confirmed by experiments BUMERANG-98 and MAXIMA-1 \cite{Jaff}.
The experimental results are consistent with the typical predictions of QG
described in the following:
the expansion of the universe and gauge theory, cosmic microwave background radiation, inflation,
candidates of dark matter, nucleosynthesis, structure formation, baryon and lepton asymmetries,
matter mass generation, $\Theta$ constant and quantum numbers, fundamental constants and cosmological parameters,
conservation laws, possible duality between the intrinsic spacetime and extrinsic spacetime,
and the relation between time and gauge boson mass.

\subsection{Universe Expansion and Gauge Theory : a New Force}

A most fundamental feature in the universe is the expansion represented by Hubble's law.
The universe expansion is first described and then a gauge theory responsible for the expansion is proposed.

\subsubsection{Universe Expansion}

Macroscopic observation for the universe expansion is Hubble's law in which the velocity of recession is given by
\begin{math}
z = H_0 D
\end{math}
where $D$ is the distance of the luminosity and $z$ is its redshift \cite{Hubb}.
$H_0 = (\Lambda_0 /3)^{1/2} \approx 10^{-42} \ \textup{GeV} \approx 10^{-28} \ \textup{cm}^{-1}$
is known as the effective Hubble constant at present, determining the expansion rate of the universe:
\begin{math}
H_0 = 100 \ h_0 \ km \ s^{-1} Mpc^{-1}
\end{math}
with $h_0 = 0.5 \sim 1$.
In terms of the line element in Robertson-Walker spacetime
\begin{equation}
d \tau^2 = dt^2 - R(t)^2 (\frac{dr^2}{1 - k r^2} + r^2 d \theta^2 + r^2 \sin^2 \theta d \varphi^2)
\end{equation}
with coordinates $(t, r, \theta, \phi)$, the scale factor $R$, and spatial curvature factor $k = 1, 0, -1$, the effective Hubble constant
is defined by
\begin{equation}
\label{huco}
\frac{\dot{R}}{R} = H_e .
\end{equation}
Note that the effective Hubble constant $H_e$ and the bare Hubble constant $H$ are manifestly distinguished in this scheme
so as to resolve several longstanding problems.
The scale factor $R$ is expanded by the power law expansion
\begin{equation}
\label{plex}
R (t) = R (0) t^{2/3}
\end{equation}
for the matter energy density $\rho_m$ with $t = 2/3H_e = 1 / (6 \pi G_N \rho_m)^{1/2}$ or the exponential expansion
\begin{equation}
\label{acex}
R (t) = R (0) \textup{exp} [\int^t_{0} H_e dt]
\end{equation}
for the vacuum density $\Lambda_e = 3 H_e^2$ with the effective Hubble constant
$H_e = (\Lambda_e /3)^{1/2} = (8 \pi G_N M_G^4/3)^{1/2} = (8 \pi G_N \langle \rho_m \rangle_e/3)^{1/2}$.

\subsubsection{Gauge Theory for a New Force}

The present expansion described above can be expressed in terms of
a gauge theory with a certain group $G'$ as a new interaction. The
massive gauge boson with $M_G^2 = M_{Pl}^2 - g_f g_g^2 \langle \phi \rangle^2 =
g_f g_g^2 [A_{0}^2 - \langle \phi \rangle^2]$ due to the vacuum energy is
responsible for repulsion, which is the source of the expansion in
the universe. The gauge boson mass reduces to the very small value
at present from the Planck mass at the Planck epoch; the extremely
small mass $M_G \approx 10^{-12}$ GeV corresponds to the nearly
zero cosmological constant $\Lambda_0 \approx 10^{-84} \
\textup{GeV}^2$, the Hubble constant $H_0 \approx 10^{-42}$ GeV,
or vacuum energy density $V_0(\bar \phi) = \langle \rho_m \rangle_0 \approx
10^{-47} \ \textup{GeV}^4$. The deceleration parameter $q_0 \equiv
- (\ddot{R}(t_0)/R(t_0))/H_e^2 = \Omega_m (1 + 3 P/\rho_m)/2$
becomes $q_0 = - \Omega_m$ for a vacuum dominated era, which
implies the accelerating expansion because of $\ddot{R} > 0$. The
effective coupling constant $G_S$ like Newton gravitation constant
$G_N$ can be similarly defined for the repulsion
leading to the accelerating expansion:
\begin{equation}
\label{neef}
\frac{G_S}{\sqrt{2}} = \frac{\sqrt{2} r_f g_r^2}{8 M_G^2} = \frac{\sqrt{\pi} r_f g_r^2 \sqrt{G_N}}{2 \sqrt{\Lambda_e}}
= \frac{\sqrt{\pi} r_f g_r^2 \sqrt{G_N}}{2 \sqrt{3} H_0}
\end{equation}
with the coupling constant $g_r$ is inversely proportional to the effective Hubble constant $H_e = H_0$ at the present epoch
and $r_f$ is a charge factor.
The gauge boson has essentially the Yukawa type potential with its extremely small mass $10^{-12}$ GeV:
the strength ratio $G_N/G_S$ is thus $10^{-61}$, which means the extremely strong,
effective coupling constant $G_S$.
In this scheme, the gauge group chain is $G \supset SU(2)_L \times U(1)_Y \times SU(3)_C \supset G'$
and the effective coupling constant hierarchy is $G_N \supset G_F \times G_R \supset G_S$.
The effective Hubble constant $H_e$ describes the accelerating expansion of (\ref{acex}) in the early universe.
The effective cosmological constant
$\Lambda_e = 8 \pi G_N M_G^4 = 8 \pi G_N \langle \rho_m \rangle_e$,
the bare vacuum energy density $\langle \rho_m \rangle = M_{Pl}^4 \approx 10^{76} \ \textup{GeV}^4$, and
the bare cosmological constant $\Lambda = 8 \pi G_N (- 2 M_{Pl}^2 g_f g_g^2 \langle \phi \rangle^2 + g_f^2 g_g^4 \langle \phi \rangle^4)$ are realized.
The repulsion has a stronger coupling constant $g_r$ than any fundamental forces known at present if $G'$ is a non-Abelian group.
Possible conserved charges are relevant for intrinsic rotational and vibrational degrees of freedom:
the intrinsic angular momentum is likely quantized in integer numbers.
Dynamics as an $SU(3)_R$ gauge theory leading to an $SU(2)_B \times U(1)_A$ gauge theory and then a $U(1)_g$
gauge theory via phase transitions might be a candidate dynamics responsible for the expansion of the universe.
The charge quantization is likely
\begin{equation}
\hat Q_g = \hat B_3 + \hat A/2
\end{equation}
where the quantum number of $B_3$ is $\pm 1/2$ and the quantum number of $A$ is $1$;
this is very analogous to the electric charge quantization $\hat Q_e = \hat I_3 + \hat Y/2$
with the third component of isospin $I_3$ and the hypercharge $Y$ in electroweak interactions.
The energy due to the quantum number $A$ is related to the intrinsic property of the massive gauge boson.
During DSSB, discrete symmetry breaking and current nonconservation are expected although they are very small.
During DSSB, non-zero mass gauge bosons and massless gauge bosons are furthermore created just as intermediate vector bosons and photons are created at
the electroweak phase transition and gluons and photons are created at the strong phase transition.
One possible source responsible for this dynamics is an intrinsic angular momentum like spin, isospin, or colorspin:
matter (or dark matter) particles possess symmetric configurations in their charge exchange for repulsion
but asymmetric configurations in their exchange for attraction.
Another possibility is the magnetic monopole with intrinsic angular momentum as one of candidates for dark matter.
The other candidate responsible for this process is the diatomic molecular rotation as baryonic matter;
when molecules rotate their energy levels are separated by roughly
$10^{-12}$ GeV and their corresponding typical wave length is $0.1$ mm.
The existence of non-zero mass gauge boson is supported by the recent BUMERANG-98 and MAXIMA-1 experiments \cite{Jaff}
and massless gauge boson is hinted by cosmic microwave background radiation (CMBR), which is more discussed in the following.

\subsection{Cosmic Microwave Background Radiation}

CMBR provides the fundamental evidence that the universe experiences the expansion via phase transition.
The radiation background is dominated by isotropic components with the thermal Planckian form
at the temperature $2.7$ K in its microwave spectrum, suggesting the radiation has almost completely relaxed to thermodynamic equilibrium.
Even though it is known that CMBR offers a firm confirmation of the hot big bang cosmology, it
can also be supporting evidence for both massless and non-zero mass gauge bosons
at the present phase transition from the viewpoint of gauge theory.
This means that CMBR is explained in terms of the ongoing astrophysical processes rather than the unobservable primordial processes.
Massless and non-zero mass gauge bosons of the gauge group responsible for the universe expansion yield a thermal radiation background
in the microwave at the present DSSB of the $SU(2)_B \times U(1)_A \rightarrow U(1)_g$ gauge group.
CMBR was first predicted by Gamow \cite{Gamo} and detected by Penzias and Wilson \cite{Penz}.
The signature of the radiation due to nearly massless and completely massless gauge bosons at phase transition is its spectrum, which is very close
to the Planckian form.
This exhibits the signature of phase transition with nearly massless gauge modes
with the characteristic mass $M_G \approx 10^{-12}$ GeV and massless gauge modes as NG bosons
with the characteristic energy $E_\gamma \approx 3 \times 10^{-13}$ GeV, which
corresponds to the typical frequency $10^{10}$ Hz and wave length $1$ mm.
Thermal equilibrium is manifest since the ratio of the interaction rate
$\Gamma \sim T^5/M_G^4$ to the expansion rate $H_e \sim T^2/M_{Pl}$ is
$\Gamma / H_e \sim T^3 M_{Pl}/M_G^4 >> 1$ at $T \simeq 2.7$ K.
CMBR is left over when the expanding universe is relaxed to thermal equilibrium, filling space with black body radiation.
In this scheme, the microwave spectrum at the temperature $2.7$ K, corresponding to the energy $3 \times 10^{-13}$ GeV, represents
the black body radiation of massless gauge bosons (photons) responsible for the expansion of the universe.
Note that the coupling constant mediated by the photon is quite different from the coupling constant $\alpha_e \approx 1/137$ mediated
by the photon in electromagnetic interactions.

The energy density of non-zero mass gauge bosons at present
is given by $\langle \rho_m \rangle_e = \rho_G = M_G^4 \approx 10^{-47} \ \textup{GeV}^4$,
their number density by $n_G \approx M_G^3 \approx 10^{-36} \ \textup{GeV}^3 \approx 1.3 \times 10^5 \ \textup{cm}^{-3}$,
and their total number by $N_G \approx 10^{91}$.
In black body radiation at temperature $T$, the number of photon
per unit volume is expressed by
\begin{equation}
\label{blra}
n_\gamma = \frac{2 \zeta (3) T^3}{\pi^2}
\end{equation}
where $\zeta$ is the Riemann zeta function.
At temperature $T = 2.7$ K, the energy density of massless gauge bosons is
$\rho_\gamma \approx 10^{-51} \ \textup{GeV}^4 \approx 4.7 \times 10^{-34} \ g \ \textup{cm}^{-3}$,
their number density is
\begin{equation}
\label{gade}
n_\gamma \approx 420 \ \textup{cm}^{-3} ,
\end{equation}
and their total number is $N_{t \gamma} \approx 10^{88}$.
Therefore, the number density of massive gauge bosons, which is evaluated by the
present Hubble constant $H_0$, is roughly $10^3$ times greater than the
number density of massless gauge bosons:
this is a strong evidence for several massive bosons, representing
non-Abelian gauge theory, compared to one massless gauge boson.

\subsection{Inflation}

The basic concept of inflation is that the effective vacuum energy in this scheme was
the dominant component of the energy density of the universe during an epoch early in
the history of the universe. Even the present epoch is a vacuum dominated time since
the vacuum energy density is bigger than the matter energy density. The present
cosmological constant $\Lambda_0$ or the Hubble constant $H_0$ is a positive value
although it is very small: the deceleration parameter is $q_0 = - \Omega_m = -
\rho_m/\rho_c$, which indicates the accelerating expansion at present. Einstein
equation \cite{Eins} provides the exponential expansion (\ref{acex}) for the effective
cosmological constant $\Lambda_e = 3 H_e^2$: $R (t) = R (0) \ \textup{exp}(\int^t_0
H_e dt)$ and $H_e = \dot R / R$ give the basic idea for the inflation scenario of the
universe evolution. This exponential expansion corresponds to the de Sitter line
element
\begin{equation}
d \tau^2 = dt^2 - e^{2 H_e t} (dr^2 + r^2 d \theta^2 + r^2 \sin^2 \theta d \varphi^2) .
\end{equation}
Since the scale factor grows exponentially during the era known as the de Sitter phase,
the radius of spatial curvature $R_c = i M_G^{-1}/(\Omega -1)^{1/2} = R(t)/|k|^{1/2}$ can grow from a small size
$R_c \approx 10^{30} M_G^{-1} \approx (10^{-11} \ \textup{GeV})^{-1} \approx 10^{-3} \ \textup{cm}$ at the Planck epoch to
the Hubble radius $H_0^{-1} \approx 10^{30} M_G^{-1} \approx (10^{-42} \ \textup{GeV})^{-1} \approx 10^{28} \ \textup{cm}$ at the present epoch so that it easily encompasses
the comoving volume that becomes the present observable universe.
In this scheme, $R_c = i H_0^{-1}/(\Omega -1)^{1/2}$ in general relativity is replaced with $R_c = i M_G^{-1}/(\Omega -1)^{1/2}$.
This also implies that the universe is always extremely flat since $\Omega -1 \approx - 10^{-61}$ or $k \approx 0$.
At this stage, the maximum wavevector mode of massive gauge bosons
$N_R \approx 10^{30}$ is introduced as a conserved quantity:
the radius of spatial curvature $R_c \approx N_R M_G^{-1}$ and the total gauge boson number $N_G = 4 \pi N_R^3/6 \approx 10^{91}$.
Note that the effective Hubble constant in the early universe is large enough to make inflation.
The possibility of a universe dominated by the effective vacuum energy is much more relevant for the realization
that the universe may undergo a series of phase transitions associated with DSSB.
Phase transition associated with DSSB offers the expansion mechanism whereby the early universe may be
dominated by the effective vacuum energy for some period of time.

Gauge theories at the Planck era and the present era justify the
above explanation of the inflation theory \cite{Guth}, which
solves the flatness problem and horizon problem with a non-zero
large cosmological constant making an exponential expansion rather
than the standard power law expansion. As expected the effective
vacuum energy density $V_e (\bar \phi) = \langle \rho_m \rangle_e \approx
M_{Pl}^4 \approx 10^{76} \ \textup{GeV}^4$ in the early Planck
universe is much higher than the effective vacuum energy density
$V_e = \langle \rho_m \rangle_e \approx 10^{-47} \ \textup{GeV}^{4}$ in the
present universe. The exponential expansion emerges whenever a
symmetry is broken spontaneously and is large enough to solve the
flatness and horizon problem since the effective cosmological
constant $\Lambda_e$ or vacuum energy density $\langle \rho_m \rangle_e$ has the
difference $10^{122}$ in the order of magnitude between at the
Planck time and at the present time. This inflation scenario is
quite reasonable from the gauge theory point of view because the
energy difference from the normal vacuum to the physical vacuum
due to the condensation of singlet gauge fields can drive the
system to expand exponentially. The gauge boson mass ratio between
at the Planck era and at the present era or the maximum wavevector
mode of gauge bosons, $N_R \approx 10^{30}$, is thus the source of
the inflation.

The inflation in this scheme easily solves the flatness and horizon problems.
At the Planck epoch $t_{Pl} \approx 10^{-43}$ s, $\Omega - 1 = - 10^{-61}$ is expected and this is known as the flatness problem.
This implies that the curvature is extremely flat.
At an epoch $t$, the proper radius of the particle horizon is expressed by $R_L = 2 t$.
The radius is extremely small, for example, $10^{-3}$ cm at the Planck epoch but is extremely large, $10^{28}$ cm, at the present epoch $t_0 \approx 10^{17}$ s:
this is the horizon problem.
The radius  $10^{-3}$ cm is the $10^{30}$ order of magnitude greater than the Planck length $l_{Pl} = M_{Pl}^{-1} \approx 10^{-33}$ cm and
the difference between two lengths originates from the maximum wavevector mode of gauge bosons
$N_R$ as a conserved quantity:
the total gauge boson number is $N_G = 4 \pi N_R^3/3 \approx 10^{91}$.
According to this scheme, the inflation during DSSB takes place and the expansion due to the vacuum energy is large enough to resolve
both the flatness and horizon problems.
Furthermore, the problem of the universe size is resolved by the maximum wavevector mode
$N_R$ since $R = N_R / M_G \approx 10^{30}/ M_G$:
the present universe size $R_0 \approx 10^{30}/ (10^{-12} \ \textup{GeV}) \approx 10^{28}$ cm.

\subsection{Candidates of Dark Matter}

This scheme may explain the reason why massive gauge bosons with the extremely
small mass $M_G \approx 10^{-12}$ GeV are plausible mediators of nonbaryonic dark matter.
Furthermore, several fermion candidates of dark matter mediated by massive gauge bosons at the present universe are considered.

The rotational behavior of galaxies indicates that the actual mass density
$\rho_m$ of the universe is much larger than the luminous mass;
this is known as the invisible dark matter problem \cite{Blom,Schr}.
Baryonic dark matter is proposed as a candidate of dark matter since the nucleosynthesis leads to
$\Omega_B \equiv \rho_B / \rho_c \approx 0.1$ with the baryon mass density
$\rho_B$ and the critical mass density $\rho_c$ \cite{Yang}.
The halo around galaxies and clusters might be considered to be a baryonic dark matter
including black hole and small massive object.
Non-baryonic matter is also strongly proposed since Euclidean metric is equivalent to
$\Omega_m = \rho_m / \rho_c \approx 1$, that is, $\rho_m \simeq \rho_c$ if the
cosmological constant is nearly zero at the present universe.
Non-baryonic dark matter must in this case be larger than the total baryonic matter
and several candidates of dark matter as weakly interacting massive particles (WIMPs) are suggested:
neutrinos, magnetic monopoles, supersymmetric particles such as gravitinos
or photinos, neutralinos, or other exotic particles \cite{Schr,Kolb}.

Massive gauge bosons responsible for the expansion of the universe can thus be considered to be strong mediators of
nonbaryonic dark matter since $\Omega_G = \rho_G/\rho_c \simeq 1$ is manifest if they are taken into account:
the energy density of gauge bosons $\langle \rho_m \rangle_e = \rho_G = M_G^4 \approx 10^{-47} \ \textup{GeV}^4$,
the number density of gauge bosons $n_G = M_G^3 \approx 10^{-36} \ \textup{GeV}^3 \approx 10^5 \ \textup{cm}^{-3}$,
the critical energy density $\rho_c = \Lambda_e/8 \pi G_N \approx 10^{-47} \ \textup{GeV}^4$, and
$\rho_G \approx \rho_c$ in this scheme.
Invisible gauge bosons with the mass $M_G \approx 10^{-3}$ eV and the number density $n_G \approx 10^5 \ \textup{cm}^{-3}$ are
analogous to invisible axions \cite{Kim}.
They are however strongly interacting massive particles (SIMPs) rather than WIMPs:
the effective coupling constant $G_S$ is about $10^{61}$ times stronger than $G_N$.

\subsubsection{Nonbaryonic Dark Matter}

A plausible candidate is nonbaryonic dark matter: lepton matter like massive neutrinos
\cite{Cows}, confirmed by the recent Super-Kamiokande experiment \cite{Fuku},
interacting with the distance of roughly $0.1$ mm. This scenario is as follows. The
photon number density (\ref{blra}) gives a neutrino density of about $330 \
\textup{cm}^{-3}$ at the time of neutrino decoupling for Majorana neutrinos. The
number of neutrinos is $10^{10}$ times greater than the number of baryons. If there
are three different flavors of light neutrinos, then the critical density $\rho_c$
implies for the mean neutrino mass $m_{\nu} \leq 17$ eV. By the fact that fermion
masses increase from generation to generation $m(\nu_e) : m(\nu_\mu) : m(\nu_\tau) =
m_e : m_\mu : m_\tau = 1: 207: 3491$, the electron neutrinos have the mass $m(\nu_e)
\leq 10^{-3}$ eV. Since electron neutrino with the mass $10^{-3}$ eV has the inertia
moment $I = 10^{-20}$ GeV $\textup{m}^2$ with the distance $0.1$ mm in this case, the
angular frequency of electron neutrino is $\omega = 10^{11}$ Hz and velocity is $v
\approx 10^7$ m/s: hot dark matter. Three generations of neutrinos might possess
intrinsic vibrational and rotational degrees of freedom as the sources of an $SU(3)$
gauge theory: this is very analogous to three color sources for the $SU(3)_C$ gauge
theory. In this scenario, several interesting aspects appear. Neutrino oscillation as
the consequence of the mass of neutrino \cite{Berg} is the analogy of quark mixing.
The strong bindings between leptons are expected; the bindings of neutrino-neutrino,
electron-neutrino, electron-electron, etc. might make numerous exotic particles. The
eightfold way of leptons might emerges as the analogy of the eightfold way of hadrons.
Electron matter is further discussed in the subject of the lepton asymmetry.

Other nonbaryonic candidates might also be considered: for
instance, they can have masses in the range from $10^{19}$ GeV to
$10^{-12}$ GeV at the Planck epoch and they can have masses in the
range from $10^{-12}$ GeV to $10^{-42}$ GeV at the present scale.
This suggests that WIMPs with high masses may be created at high
energies as the candidate of invisible dark matter but SIMPs with
low masses may be created at low energies as the candidates of
invisible dark matter.

\subsubsection{Monopole}

Some comments on electric and magnetic monopoles as candidates of dark matter
are available from the QG point of view.

Magnetic monopole is another fermion candidate of dark matter.
Gravitational magnetic monopole, isospin magnetic monopole, and
color magnetic monopole may be considered. Matter particles are
created as gravitational electric monopoles at the Planck epoch
and the some parts of gravitational magnetic monopoles disappear
due to the breaking of discrete symmetries. Even at lower
temperatures, isospin magnetic monopoles at the weak scale or
color magnetic monopoles at the strong scale are not allowed
because of the violation of discrete symmetries as expected by
observation. They have the odd parity, which is not observed in
measurement, when the parity is considered, for example. According
to $\Theta$ values, $\Theta_{Pl} = 10^{61}$ at the Planck epoch
reflects the electric monopole dominant matter over the magnetic
monopole dominant vacuum and $\Theta_{EW} = 10^{-4}$ at the weak
epoch and $\Theta_{QCD} = 10^{-10}$ at the strong epoch also reflect
the electric monopole dominant matter as visible matter even
though the asymmetry of discrete symmetries is less than that at
the Planck epoch.

Electric monopole at low energies may be considered as the candidate of dark matter
from the viewpoint of the possible gauge theory. Electric monopole at low energies is
likely a spin $1/2$ particle with the strong coupling constant, comparable to the
magnetic coupling constant $g_m = 69 \ e$ explained by Dirac quantization $e g_m =
2 \pi n$ \cite{Dira}, and with the extremely light mass rather than heavy mass. It may be
an extended object \cite{Hoof3} less than a cutoff scale ($\sim 10^{-3}$ cm) but a
pointlike object in the universe. The effective coupling ($G_S$) at low energy is also
very strong. Note that the charge quantization is likely $\hat Q_g = \hat B_3 + \hat
A/2$, which is very analogous to the electric charge quantization $\hat Q_e = \hat I_3
+ \hat Y/2$ in electroweak interactions. This suggests that electric monopole may be
postulated as a spinor with the angular momentum $l=0$ and the even parity and have
the very strong coupling constant. The gauge boson identified as the electric string
connecting two electric monopoles has the mass around $10^{-12}$ GeV corresponding to
the distance $10^{-3}$ cm and its number density is around $n_m \sim n_G \approx 10^5 \
\textup{cm}^{-3}$, which is bigger than the number density of massless gauge bosons
$n_\gamma \approx 420 \ \textup{cm}^{-3}$. The electric monopole might have the mass
$m_f \approx 10^{-5}$ eV for the strong coupling constant and the fermion even-odd
parity singlet difference number $N_{sd} =1$ when $M_G \approx 10^{-12}$ GeV from the
relation $M_G = \sqrt{\pi} m_f g_f \alpha_g \sqrt{N_{sd}}$ based on the following subsection: $m_f
\approx 10^{-2}$ eV when $M_G \approx 10^{-9}$ GeV.

The pairing between electrons, also relevant for the pairing between magnetic monopoles in the vacuum when the electric-magnetic duality is applied, might be the origin of
high temperature superconductivity and the mechanism may be understood in terms of a gauge theory
as the dual Meissner effect, which is based on the perfect dielectric in the vacuum:
this is highly speculative from the viewpoint of the electric-magnetic duality
at a high critical temperature like $300$ K,
for example, since the binding energy of the Cooper pair is about $10^{-3}$ eV and the distance between electrons of the Cooper pair is about $10^{-6}$ m.
Photon or phonon attraction due to electrons or ions at low energies might be the source of pairing mechanism for high $T_c$ superconductivity.

\subsection{Nucleosynthesis}

Primordial nucleosynthesis may also be a good probe to test this
scheme. Nuclear reactions result in the production of substantial
amounts of D, ${^3}\textup{He}, \ {^4}\textup{He}$, and
${^7}\textup{Li}$ when they took place from $t \simeq 0.01$ sec to
$10^2$ sec or $T \simeq 10$ MeV to 0.1 MeV. The observed abundance
data are $\textup{D/H} \simeq 10^{-5}$,
${^3}\textup{He}/\textup{H} \simeq 10^{-5}$,
${^4}\textup{He}/\textup{H} \simeq 0.25$, and ${^7}\textup{Li/H}
\simeq 10^{-10}$. The deuterium and helium-4 are particularly
important since there are no contemporary astrophysical processes
that can explain their observed abundance. The abundance is thus
considered to be relics of primordial nucleosynthesis as a result
of nuclear interactions around nuclear energies $10 \sim 0.1$ MeV.
Nucleosynthesis may be explained in terms of quantum
nucleardynamics (QND) \cite{Roh3,Roh31} as an $SU(2)_N \times U(1)_Z$
gauge theory originated from QCD as an
$SU(3)_C$ gauge theory. The essence of this approach is the role
of massive gauge bosons, gluons, with the mass around $300$ MeV
and the role of massless gauge bosons, photons, as NG bosons.
Primordial nucleosynthesis leading to the abundance of the light
particles may be calculated since parameters of strong
interactions as well as weak interactions are known.

The ratio of neutrons and protons is of particular importance to the outcome of primordial nucleosynthesis, as essentially
all neutrons in the universe become incorporated into ${^4}\textup{He}$.
The ratio of neutrons to protons is uniquely determined at the time nucleosynthesis begins, once parameters of weak interactions process are known
since the balance between neutrons and protons is maintained by weak interactions \cite{Glas}.
The ratio is found to be
\begin{equation}
\frac{N_n}{N_p} = \textup{exp}(\frac{m_p - m_n}{T}) \simeq \textup{exp}(- \frac{1.29}{T_{10}})
\end{equation}
where $T_{10}$ indicates the temperature expressed in units of $10^{10}$ K.
The ratios of neutrons to protons thus become $1:1$ at $T \geq 10^{12}$ K, $5:6$ at $T = 10^{11}$ K, $3:5$ at $T = 3 \times 10^{10}$ K.
The ratio is $1/6$ at $10^{9}$ K, where the primordial synthesis of ${^4}\textup{He}$ begins.
At this point, the deuterium abundance grows large enough for the deuterons to burn to the helium.

The sensitivity of the abundance depends on a cosmological
parameter, the baryon number asymmetry $\delta_B$, and two
physical parameters, the total number of effective massless
degrees of freedom $g_*$ and the neutron half lifetime $\tau_n$
\cite{Kolb}. Since the weak interaction rate $\Gamma_w \propto
T^5/\tau_n$ provides the freeze-out energy $T_f \propto
\tau_n^{1/2}$, the increase in the neutron half lifetime leads to
the increase of the n/p value and then leads to the increase of
the abundance. An increase in $g_*$ makes a faster expansion
rate and also makes an early freeze-out of the $n/p$ value at
higher energy since $T_f \propto g_*^{1/6}$. The photon degree of
freedom as the NG boson, which appears as the result of DSSB of
QCD, should be included to $g_*$ value. The abundance is
proportional to $\delta_B^{A-1}$ with the baryon asymmetry
$\delta_B$ and the nuclear mass number A, which are governed by
QND \cite{Roh3,Roh31}. For a larger value of $\delta_B$, the abundance
of light particles starts earlier and thus ${^4}\textup{He}$
synthesis occurs earlier, when the n/p ratio is larger, resulting
in more ${^4}\textup{He}$. The sensitivity to $\delta_B$ is much
more significant with the yields of D and ${^3}\textup{He}$
decreasing as $\delta_B^{-i}$ with $i \simeq 1 \sim 2$.

Primordial nucleosynthesis provides the most precise determination
of the baryon density; the baryon number asymmetry $\delta_B =
(N_B - N_{\bar B})/ N_{t \gamma} \approx 10^{-10}$ is observed in
the present universe as discussed in the following subsection.
Primordial nucleosynthesis implies the fraction of the critical
density in baryons, $\Omega_B$, must be less than one. If
$\Omega_m \approx \rho_m/\rho_c$ is equal to one, primordial
nucleosynthesis offers the strong indication that the most form of
the mass density of the universe is in a form other than baryons.
Primordial nucleosynthesis needs to be more precisely investigated
for fine tuning since QCD \cite{Frit}, QND \cite{Roh3,Roh31},
QWD \cite{Roh2}, GWS model \cite{Glas}, and QG \cite{Roh1} as underlying gauge
theories are now at hand.

\subsection{Structure Formation}

This scheme is consistent with the observable data for the structure formation of the universe.
The structure formation is understood by a calculable relationship
\begin{equation}
\frac{\delta \rho_m}{\rho_m} \propto \frac{\delta T}{T}
\end{equation}
where $\rho_m$ is the matter energy density.
According to the data of CMBR, $\delta T/T \leq 10^{-4}$ on angular scales ranging from $1$ arc second to $180^\circ$ is obtained.
The data of CMBR mean that the present universe is very isotropic and
homogeneous in the large scale.
The value of fluctuation is not enough to explain the structure formation of the universe.
However, if the dark matter candidate described above is taken into account, the structure formation might be explained;
for instance, the structure formation through massive neutrinos \cite{Bond}.
The presence of dark matter plays a significant role in the formation of structure since this allows a large
$\delta \rho _m/ \rho_m$ of nonbaryonic fluctuation at the recombination era, which might match to the period of DSSB
from the $SU(3)_R$ to the $SU(2)_B \times U(1)_A$ gauge theory.
Then the baryonic fluctuation, which was small at that epoch, catches up with the large fluctuations of the nonbaryonic matter at the later epoch
since the two kinds of matter interact strongly.
Massive gauge bosons considered as the candidates of dark matter in this scheme produce large fluctuation and strongly interact
at the matter-dominated epoch so as to form the large scale of the universe:
the vacuum energy density is bigger than the matter energy density at present
as verified by experiments BUMERANG-98 and MAXIMA-1 \cite{Jaff}.

In order to understand the formation of structure precisely, it is important to know the initial conditions at the time structure formation began.
The formation of matter structure began where the universe became matter-dominated around $t \approx 4.4 \times 10^{10}$ s or
$E \approx 5.5$ eV, when density perturbations in the matter component of the universe began to grow.
The time of matter-radiation equality might be the initial epoch for matter structure formation.
The initial data present that the total amount of nonrelativistic matter in the universe is quantified by
$\Omega_m = \rho_m/\rho_c \approx 1$ and
the fraction of baryon is estimated by $\Omega_B \approx 0.1$ while the fraction of dark matter is estimated by
$\Omega_D > \Omega_B$.
At the epoch of matter-radiation equality gauge bosons have the mass $M_G \sim 5.5 $ eV
with the particle number density $n_G \approx 10^{15} \ \textup{cm}^{-3}$ and
form atoms, whose size extends to the galaxy size $10^{30}$ times greater than an atom size:
the large structure might be formed simultaneously.
A detailed scenario of structure evolution may be constructed by numerical
simulation and the result can be compared to the universe observed today.

In order to estimate the fluctuations including strongly interacting dark matter, thermodynamic equilibrium for the system is considered.
In thermodynamic equilibrium, the fluctuation of internal energy $\delta U$ is given by
\begin{math}
\delta U = T C_V^{1/2}
\end{math}
which leads to
\begin{math}
\frac{\delta U}{U} = C_V^{1/2}
\end{math}
where $C_V$ is the heat capacity.
Since the heat capacity $C_V$ grows linearly with the size of the system,
the fractional energy fluctuations $\delta U/U$ fall as the square root of the system size.
Specially, the heat capacity diverges during phase transition.
The fluctuation of DSSB at the time of matter-radiation equality is thus large enough to form the large
scale structure and becomes small as the system expands.
This is consistent with the present, small fluctuation $\delta T/T \leq 10^{-4}$.
Therefore, the probable fluctuation at the time of matter-radiation equality
\begin{math}
\frac{\delta \rho_m}{\rho_m} \leq 0.01
\end{math}
is deduced from the fluctuation $\delta T/T \leq 10^{-4}$ from the data of CMBR.
In addition to the large fluctuation, the very strong interaction of dark
matter due to the very light gauge bosons $10^{-12}$ GeV expedites the
larger structure formation of the universe.
Note that the vacuum energy is greater than the baryon matter energy.

For example, in the solar system, the sun has the mass $\sim
10^{57}$ GeV, the mass density $\sim 5 \ \textup{g \ cm}^{-3}$,
and the radius $\sim 10^{10}$ cm and the earth has the mass $\sim
10^{51}$ GeV, the mass density $\sim 5 \ \textup{g \ cm}^{-3}$,
the radius $\sim 10^{8}$ cm, and the distance from the sun $\sim
10^{12}$ cm. This suggests that the formation of the solar system
might be initiated at the atomic scale $\sim 10^{-8}$ cm according
to the baryon matter density $\sim 5 \ \textup{g \ cm}^{-3}$ in
the matter space at the atomic scale. Note that the total baryon
number $10^{78}$ is the same order of the total photon number
$10^{78}$ in the bound state of nucleons-electrons at the atomic
scale: this is the hint of the matter-radiation equality epoch
around the atomic epoch. Similarly, the galaxy with the typical
size $\sim 10^{22}$ cm with the mass density $10^{-24} \ \textup{g
\ cm}^{-3}$ might be also formed at the atomic scale since the
mass density corresponds to one at the atomic scale when the
vacuum volume is used because baryons and gauge bosons are
respectively quantized by $N_F \simeq 10^{26}$ and $N_R \simeq
10^{30}$: refer to Table \ref{fuco} discussed later. Because the
predicted baryon number density is $n_B = 10^{-31} \ \textup{g \
cm}^{-3}$ in the universe, the mass density of the galaxy is
$10^{7}$ times greater than the baryon number density of the
universe.

\subsection{Baryon Asymmetry and Lepton Asymmetry}

The baryon asymmetry \cite{Stei0} is apparent in the present universe.
Baryogenesis due to gravity are at present very weak by the effective coupling constant $G_N \approx 10^{-38} \ \textup{GeV}^{-2}$.
Although equal quantities of matter and antimatter at the Planck scale were expected, the universe is presently quite asymmetric in their ratio.
The baryon asymmetry is addressed and then the fermion asymmetry and lepton asymmetry are suggested.

\subsubsection{Baryon Asymmetry}

In terms of the baryon  energy density $\rho_B \approx 1.88 \times 10^{-29} \ \Omega_B h_0^2 \ \textup{g \ cm}^{-3}$,
the number of protons per unit volume is
\begin{equation}
\label{banu}
n_B = \rho_B/m_p \sim 1.13 \times 10^{-5} \ \Omega_B h_0^2 \ \textup{cm}^{-3} .
\end{equation}
The baryon-antibaryon asymmetry at present is, from (\ref{gade}) and (\ref{banu}), estimated by the
number of baryons dominating over the number of antibaryon by a tiny factor of $10^{-10}$ if $\Omega_B \approx 0.1$:
\begin{equation}
\label{baas}
\delta_B = \frac{N_B - N_{\bar B}}{N_B + N_{\bar B}}= \frac{N_B}{N_{t \gamma}}
\approx \frac{10^{78}}{10^{88}} = 10^{-10}
\end{equation}
where $N_{t \gamma}$ is the total number of massless gauge bosons (photons).
In order to explain the baryon asymmetry, three features are required \cite{Sakh}:
baryon number symmetry violation at the origin of time, C and CP symmetry violation, and
nonequilibrium state during C and CP violating processes.

The baryon asymmetry described above may be discussed from the gauge theory point of view.
Massive gravitons might violate discrete symmetries and the antibaryon (or antimatter) number conservation
just as the Higgs mechanism in electroweak interactions violates discrete symmetries and chiral symmetry.
The possibility implies that the baryon current anomaly triggered by DSSB
leads to the baryon number asymmetry;
the current density $J^\mu$ seems to violate the baryon number symmetry due to a current anomaly in (\ref{thet}).
The discrete symmetries of P, C, CP, and T are explicitly broken
during DSSB as the requirement of the baryon asymmetry.
If the antibaryon number current is not conserved, some antibaryonic particle spectra must disappear;
this might be a considerable explanation to the baryon asymmetry (\ref{baas}).
At the Planck temperature on the order of $T \approx M_{Pl}$, gravitons existed
in thermal equilibrium and the antibaryon current anomaly could probably generate a baryon asymmetry.
However, as the temperature of the universe decreased,
gravitons were no longer in thermal equilibrium and the baryon asymmetry was frozen permanently.
Since the interaction rate is given by $\Gamma \sim n \sigma |v| \sim G_N^2 T^5$ and the expansion rate is given by $H_e \sim T^2/M_{Pl}$,
the ratio of the interaction rate to the expansion rate becomes
$\Gamma/H_e \sim T^3/M_{Pl}^3$, which indicates nonequilibrium starts at the temperature $T\sim M_{Pl}$.
The population of gravitons and the number of antibaryons were suppressed by
the Boltzmann factor $\textup{exp} (-M_{Pl}/T)$ since $T< M_{Pl}$.
Since the population difference of baryon and antibaryon ($N_B \approx 10^{78}$) is represented in the mass
of baryon $0.94$ GeV, which is $10^{12}$ times greater than the present gauge boson mass $10^{-12}$ GeV with the particle number $n_G \approx 10^{91}$,
the population difference of matter and antimatter at present would be $N_f \simeq 10^{91}$ in the unit of mass $10^{-12}$ GeV:
\begin{equation}
\label{feas}
\delta_f = \frac{N_f - N_{\bar f}}{N_f + N_{\bar f}}= \frac{N_f}{N_{t \gamma}}
\approx \frac{10^{91}}{10^{88}} = 10^{3} .
\end{equation}
This implies that the fermion number $N_f \simeq 10^{91}$ with the
fermion mass $10^{-12}$ GeV, which possesses the baryon number
$N_B \simeq 10^{-12}$ in the unit of $1$ GeV, might be a good
quantum number just as the baryon number $10^{78}$ is a good
quantum number in the universe. The Cooper pairing mechanism of
matter particles produces the baryon asymmetry through DSSB during
the evolution of the universe. In the minimal GUT of the $SU(5)$
gauge theory \cite{Geor}, the baryon asymmetry term involved in
perturbation theory is too small to explain the observed baryon
asymmetry. Necessary ingredients mentioned above are however
explicitly satisfied through nonperturbative processes, at the
tree level, during DSSB  by QG as a gauge theory and this
indicates that the baryon asymmetry is carried by the current
anomaly and the singlet gauge boson condensation. The baryon number is not
conserved above the strong scale but the baryon number is
conserved below the strong scale as illustrated by the $U(1)_Z$
gauge theory at the strong scale \cite{Roh3}. The baryon asymmetry
is also related to both the Avogadro's number of atoms and the
nuclear matter density.

\subsubsection{Lepton Asymmetry}

The lepton-antilepton asymmetry \cite{Wein1}, which implies the lepton number
violation observable at present, is an analog of the baryon asymmetry as the
consequence of C, T, and CP violation during DSSB:
\begin{equation}
\label{leas}
\delta_L = \frac{N_L - N_{\bar L}}{N_L + N_{\bar L}} = \frac{N_L}{N_{t \gamma}}
\end{equation}
with the total lepton number $N_L = L$.
Leptogenesis due to gravity are also at present very weak by the effective coupling constant
$G_N \approx 10^{-38} \ \textup{GeV}^{-2}$ but it was very strong at the Planck epoch.
It consist of the electron asymmetry
$\delta_e = \frac{N_e - N_{\bar e}}{N_{t \gamma}} \approx \frac{10^{81}}{10^{88}} = 10^{-7}$
and the neutrino asymmetry $\delta_\nu = \frac{N_\nu - N_{\bar \nu}}{N_{t \gamma}} \approx \frac{10^{91}}{10^{88}} = 10^3$
according to the electron mass $0.5$ MeV and the probable neutrino mass around $10^{-3}$ eV under the assumption of $\Omega_e = \rho_e/\rho_c \approx 1$.
The neutrino asymmetry is $\delta_\nu = N_\nu/N_{t \gamma} \approx 10^{88}/10^{88} = 1$
if the neutrino mass $m_\nu \approx 1$ eV.
Lepton asymmetries of muon and tau
$\delta_\mu = N_\mu/N_{t \gamma} \approx 10^{79}/10^{88} = 10^{-9}$
and $\delta_\tau = N_\tau/N_{t \gamma} \approx 10^{78}/10^{88} = 10^{-10}$ are as well expected if lepton matter has
the same order with the critical density $\rho_c$.
The neutrino asymmetry $\delta_\nu \approx 10^3$ is quite reasonable if neutrinos are identified as dark matter.

These asymmetries predicted under the assumption of $\Omega_L = \rho_L/\rho_c \simeq 1$ seem to indicate the nonconservation of the lepton quantum number,
$\delta_L = \delta_B + \delta_{(B-L)}$, if the $(B - L)$ quantum number is not conserved.
Since $\delta_L = \delta_B \approx 10^{-10}$ if $B$, $L$, and $(B - L)$ quantum numbers are separately conserved below the weak energy
from the apparent neutrality of the universe and the Avogadro's number of atoms,
lepton matter as dark matter is also suggested.
This is consistent with the total photon number $10^{78}$ in the formation of atoms by binding nucleons and electrons.
The lepton number is not conserved above the weak scale but
the lepton number is conserved below the weak scale as illustrated by the $U(1)_Y$ gauge theory in weak interactions.
On the other hand, the gauge boson asymmetry is estimated by
\begin{equation}
\delta_G = \frac{N_G}{N_{t \gamma}} \approx \frac{10^{91}}{10^{88}} = 10^3 ,
\end{equation}
which also indicates the nonconservation of the $(B - L)$ quantum number but the conservation of the fermion quantum number $N_f \approx 10^{91}$
at the very high temperatures.

\subsection{Matter Mass Generation}

Conventional mass term in the Lagrangian density is not allowed but mass can be generated by the $\Theta$ term,
which might violate discrete symmetries and antimatter current conservation, as described in the following subsection.
Basically, the mass is generated through DSSB caused by the surface effect, which quantizes spacetime.
In this subsection, constituent fermions and mass formation mechanism as dual Meissner effect are proposed.

\subsubsection{Dual Meissner Effect}

The binding fermion formation is the consequence of the gravitational electric
interaction due to the dual Meissner effect, in which the gravitational electric
monopole and magnetic dipole are confined inside the fermion while the gravitational
magnetic monopole and electric dipole are confined in the vacuum. The difference
number of even-odd parity singlet fermions $N_{sd} = N_{ss} - N_{sc}$
with the even parity singlet number of constituent particles $N_{ss}$ and
the odd parity condensation number $N_{sc}$ are introduced.

The relation between the masses of the massive gauge boson and of the associated fermion can be obtained
as the dielectric mechanism \cite{Roh3} in terms of the analogy of the diamagnetism
mechanism in superconductivity \cite{Aitc}.
During the DSSB of gauge symmetry and chiral symmetry, the dual Meissner
effect of the gravitational electric field in the relativistic case can be expressed by
\begin{equation}
\label{wame0}
\partial_\mu \partial^\mu A^\mu = - M_G^2 A^\mu
\end{equation}
where the right hand side is the screening current density, $j^\mu_{sc} = - M_G^2 A^\mu$.
The dual Meissner effect of the color electric field in the
static limit is expressed by
\begin{equation}
\label{wame}
\nabla^2 \vec E_g = M_G^2 \vec E_g
\end{equation}
which shows the gravitational electric field $\vec E_g$ excluded in the vacuum by
$\vec E_g = \vec E_{g0} e^{- M_G r}$.
Note the difference between the gravitational dielectric due to the
gravitational electric field $\vec E_g$ and the
gravitational diamagnetism due to the gravitational magnetic field $\vec B_g$.
The mechanism is analogously connected with Faraday induction law, which opposes the change in the gravitational electric flux rather than
the gravitational magnetic flux, according to Lenz's law.

The massive gauge boson can be linked to the fermion mass $m_f$:
\begin{equation}
\label{gafe}
M_G = (\frac{g_{gm}^2 |\psi(0)|^2}{m_f})^{1/2} \simeq \sqrt{\pi} m_f g_f \alpha_g \sqrt{N_{sd}}
\end{equation}
where $g_{gm} = 2 \pi n/\sqrt{g_f} g_g = 2 \pi \sqrt{N_{sd}}/\sqrt{g_f} g_g$ is the gravitational
magnetic coupling constant and $|\psi(0)|^2$ is the particle
probability density. The relation is obtained by the analogy of
electric superconductivity \cite{Aitc}, $M^2 = q^2 |\psi(0)|^2/m$:
$q= - 2 e$ and $m = 2 m_e$ are replaced with $g_{gm}$ and $m_f$.
The difference number of even-odd parity singlet fermions is
$N_{sd}$, the wave function at the origin is $\psi (0)$, and the
average system size is $l \simeq 1/|\psi (0)|^{2/3} \simeq 1/m_f
g_f \alpha_g$ \cite{Roh3}. For example, $N_{sd} \simeq
10^{61}$ for the gravitational charge in the case of the fermion
with the mass $10^{-12}$ GeV, $N_{sd} \simeq 10^{12}$ for the
isospin charge in the case of the electron with the mass $0.5$
MeV, and $N_{sd} \simeq 1$ for the color charge in the case of the
proton with the mass $0.94$ GeV. Fermion mass generation mechanism
is the dual pairing mechanism of constituent fermions, which makes
bosonlike particles of paired fermions. According to the
electric-magnetic duality \cite{Dira,Hoof3,Mand,Seib}, the
gravitational electric flux is quantized by $\Phi_E = \oint \vec
E_g \cdot d \vec A = \sqrt{g_f} g_g$ in the matter space while the
gravitational magnetic flux is quantized by $\Phi_B = \oint \vec
B_g \cdot d \vec A = g_{gm}$ with the gravitational magnetic
coupling constant $g_{gm}$ in the vacuum space: the Dirac
quantization condition $\sqrt{g_f} g_g g_{gm} = 2 \pi n$
\cite{Dira} is satisfied and $n = \sqrt{N_{sd}}$ is realized. In
the matter space, it is the pairing mechanism of gravitational
electric monopoles while in the vacuum space, it is the pairing
mechanism of gravitational magnetic monopoles according to the
duality between electricity and magnetism: gravitational electric
monopole pairing and gravitational magnetic monopole condensation.
In the dual pairing mechanism, discrete symmetries P, C, T, and CP
are dynamically broken. Gravitational electric monopole,
gravitational magnetic dipole, and gravitational electric
quadrupole remain in the matter space but gravitational magnetic
monopole, gravitational electric dipole, and gravitational
magnetic quadrupole condense in the vacuum space as the
consequence of P violation. Antimatter condenses in the vacuum
space while matter remains in the matter space as the consequence
of C violation: the matter-antimatter asymmetry. The electric
dipole moment of the neutron and the decay of the neutral kaon
decay are the typical observations for T or CP violation.

\subsubsection{Constituent Fermions}

The $\Theta$ term at the Planck scale causes the DSSB and accordingly generates matter mass,
which is related to the vacuum energy represented by the gauge boson mass.
Fermions known as elementary particles are thus postulated as composite particles
consisted of constituent particles.

The relation between the gauge boson mass and the free fermion mass,
which is confirmed by (\ref{gafe}) is given by
\begin{equation}
\label{gafe0}
M_{Pl} = \sqrt{\pi} m_f g_f \alpha_g \sqrt{N_{ss}}
\end{equation}
or
\begin{math}
\label{gafe5}
M_G = \sqrt{\pi} m_f g_f \alpha_g \sqrt{N_{sd}}
\end{math}
where $N_{ss}$ is the number of singlet fermions and $N_{sd}$ is the difference number
of even-odd parity singlet fermions. Fermion mass as the result of the dual pairing
mechanism described above is composed of constituent particles:
\begin{equation}
\label{gafe1}
m_f = \sum_i^N m_i
\end{equation}
where $m_i$ is the constituent particle mass.
In the above, $N$ depends on the intrinsic quantum number of constituent particles:
$N = N_{sd}^{3/2}$.
For examples, $N = 1/B$ with the baryon quantum number $B$ for a constituent quark in the formation of a baryon,
$N = 1/M$ with the meson quantum number $M$ for constituent quark in the formation of a meson,
and $N = 1/L$ with the lepton quantum number $L$ for a constituent particle in the formation of a lepton.
The minimum mass of a fermion at the Planck scale is the $10^{-12}$ GeV for the difference number
$N_{sd} \simeq 10^{61}$, which has the intrinsic baryon number $B \simeq 10^{-12}$
or the intrinsic lepton number $L \simeq 10^{-9}$.

The difference number of even-odd parity singlet fermions $N_{sd}$ in fermion mass
generation $M_G = \sqrt{\pi} m_f g_f \alpha_g \sqrt{N_{sd}}$ thus represents $N_{sd} = N_{ss} -
N_{sc}$ where $N_{ss}$ is the number of even parity singlet fermions and $N_{sc}$ is
the condensed number of odd-parity singlet fermions. At the phase transition, $N_{sc}$
becomes zero so that $N_{sd}$ becomes the maximum. Using relations $M_G = m_f g_f
g_g^2 \sqrt{N_{sd}}$ and $M^2_G = M_{Pl}^2 - g_f g_g^2 \langle \phi \rangle^2 = g_f g_g^2 [A_{0}^2 -
\langle \phi \rangle^2]$, the zero point energy $M_{Pl}^2 = \pi m_f^2 g_f^2 \alpha_g^2 N_{ss}$ and the
reduction of the zero-point energy $\langle \phi \rangle^2 = m_f^2 g_f \alpha_g N_{sc}/4$ are obtained. There is
the condensation process in fermion mass generation mechanism. The difference number
of fermions $N_{sd}$ is the origin of symmetry violation during DSSB.  Fermions with
odd parity condense in the vacuum space while fermions with even parity remain in the
matter space; for example, magnetic monopoles with odd parity are not observed but
electric monopoles are observed in the matter space. Discrete symmetries are violated
so as to have complex scattering amplitude and the nonconservation of the charge
singlet current. This is the main reason of the change of the fermion mass and gauge
boson mass.

\subsection{$\Theta$, $\Omega$ Constants and Quantum Numbers}

The $\Theta$ vacuum term as the nonperturbative one is taken into
account to show DSSB in analogy with the axial current anomaly in
strong interactions \cite{Adle}, which is linked to the $\Theta$
vacuum in QCD as a gauge theory \cite{Hoof2}. The $\Theta$ term
representing the surface effect takes the difference in quantum
numbers of left- and right-handed fermions to generate fermion
masses. This approach thus uses the DSSB of local gauge symmetry
and global discrete symmetries to generate fermion masses. Due to
the $\Theta$ term, the boundary condition of the system is imposed
and the matter and vacuum space is quantized. This suggests that
the effect of the $\Theta$ term is not negligible even in the
present universe scale as well as in the Planck scale; of course,
the dynamical magnitude of the parameter $\Theta$ changes in the
order $10^{122}$ from the Planck epoch to the present epoch.

The parameter $\Theta$ is constrained to hold the flat universe
condition $\Omega - 1 = - 10^{-61}$ and it changes from $10^{61}$
at the Planck scale to $10^{-61}$ at the present scale via
$10^{-14}$ at the strong scale. The gauge invariance and boundary
condition of spacetime provide the quantization of the internal
and external space. Relation between $\Theta$ and $\Omega$ constants is evaluated,
the $\Theta$ constant is
related to CP violation processes such as the decay of the neutral
kaon in weak interactions and the electric dipole moment of the
neutron in strong interactions, and the $\Theta$ constant is
applied to evaluate intrinsic and extrinsic quantum numbers.

\subsubsection{$\Theta$ Constant and $\Omega$ Constant}

Under the constraint of the extremely flat universe,
which is required by quantum gauge theory and inflation scenario and is
verified by recent experiments \cite{Jaff},
the relation $\Omega - 1 = - 10^{-61}$ leads to
\begin{equation}
\Omega = (\langle \rho_m \rangle - \Theta \ \rho_m)/\rho_G = 1 - 10^{-61},
\end{equation}
where $\rho_m$ is the matter energy density, $\langle \rho_m \rangle$ is the zero point energy density, and
$\rho_G$ is the vacuum energy density.
This means that the ratio of the zero point energy density to the vacuum energy density is
\begin{math}
\langle \rho_m \rangle/\rho_G = 1
\end{math}
and the $\Theta$ constant is obtained by
\begin{math}
\Theta = 10^{-61} \ \rho_G/\rho_m.
\end{math}
If the matter energy density in the universe is $\rho_m \simeq \rho_c \simeq 10^{-47} \ \textup{GeV}^4$ and is conserved,
the $\Theta$ constant in (\ref{thet}) depends on the gauge boson mass $M_G$ since $\rho_G = M_G^4$:
\begin{equation}
\label{thvu}
\Theta = 10^{-61} \ M_G^4/\rho_c .
\end{equation}
$\Theta$ values becomes $\Theta_{Pl} \approx 10^{61}$ at the Planck scale,
$\Theta_{EW} \approx 10^{-4}$ at the weak scale, $\Theta_{QCD} \approx
10^{-12}$ at the strong scale,
and $\Theta_{0} \approx 10^{-61}$ at the present scale.
This is consistent with the observed results, $\Theta < 10^{-9}$ in the
electric dipole moment of the neutron \cite{Alta} and $\Theta
\simeq 10^{-3}$ in the neutral kaon decay \cite{Chri} as CP
violation parameters.
Since the weak boson mass changes by the Weinberg mixing angle $\theta_W$, $M_G \rightarrow M_G \sin \theta_W$,
during the DSSB of $SU(2)_L \times U(1)_Y$ or $SU(2)_L \times U(1)_Y \rightarrow U(1)_e$ gauge theory,
the change of the $\Theta$ constant in electroweak interactions depends on $\theta_W$:
\begin{equation}
\label{thvr1}
\Delta \Theta \propto \sin^4 \theta_W = i_f^{e 2}.
\end{equation}
Similarly, since the gluon mass changes, $M_G \rightarrow M_G \sin \theta_R$,
during the DSSB of $SU(3)_C \rightarrow SU(2)_N \times U(1)_Z$ or $SU(2)_N \times U(1)_Z \rightarrow U(1)_f$ gauge theory,
the change of the $\Theta$ constant in strong interactions depends on the color mixing angle $\theta_R$:
\begin{equation}
\label{thvr2}
\Delta \Theta \propto \sin^4 \theta_R = c_f^{f 2}.
\end{equation}
Note that the isospin factor $i_f^n = \sin^2 \theta_W \simeq 1/4$, the weak boson mass $M_W = M_Z \cos \theta_W$,
the color factor $c_f^n = \sin^2 \theta_R \simeq 1/4$, and the gluon mass $M_A = M_B \cos \theta_R$.
The relation between the $\Theta$ constant and the difference number $N_{sd}$ is given by
\begin{equation}
\Theta = \pi^2 m_f^4 g_f^4 \alpha_g^4 N_{sd}^2/10^{61} \rho_c
\end{equation}
from equations (\ref{thvu}) and (\ref{gafe5}).

\subsubsection{$\Theta$ Constant and Quantum Numbers}

The invariance of gauge transformation provides
$\psi [\hat O_\nu] = e^{i \nu \Theta} \psi [\hat O]$ for the fermion wave function $\psi$ with the transformation of an operator $\hat O$ by
the class $\nu$ gauge transformation, $\hat O_\nu$:
the vacuum state characterized by the constant $\Theta$ is called the $\Theta$ vacuum \cite{Hoof2}.
The true vacuum is the superposition of all the $|\nu \rangle$ vacua with the phase $e^{i \nu \Theta}$:
$|\Theta \rangle = \sum_\nu e^{i \nu \Theta} |\nu \rangle$.
The topological winding number $\nu$ or the topological charge $q_s$ is defined by
\begin{equation}
\label{tonu}
\nu = \nu_+ - \nu_- = \int \frac{g_f g_g^2}{16 \pi^2} Tr G^{\mu \nu} \tilde G_{\mu \nu} d^4 x
\end{equation}
where the subscripts $+$ and $-$ denote moving particles with opposite
intrinsic properties in the presence of the gauge fields \cite{Atiy}.
The subscripts $+$ and $-$ respectively represent antimatter and matter
particles at the Planck scale, right-handed and left-handed particles at
the weak scale, axial-vector and vector particles at the strong scale.
The matter energy density generated by the surface effect is postulated by
\begin{equation}
\label{toma}
\rho_m \simeq \rho_c \simeq \frac{g_f g_g^2}{16 \pi^2} Tr G^{\mu
\nu} \tilde G_{\mu \nu} \simeq 10^{-47} \ \textup{GeV}^4
\end{equation}
which implies that the fermion mass is generated by the difference of fermion numbers
moving to backward and forward directions at the Planck scale. The difference number
$N_{sd}$, the singlet fermion number $N_{ss}$, and the condensed singlet fermion
number $N_{sc}$ in intrinsic two-space dimensions respectively correspond to $\nu$,
$\nu_+$, and $\nu_-$ in three-space and one-time dimensions.  In the presence of the $\Theta$
term, the odd singlet current is not conserved due to an Adler-Bell-Jackiw anomaly
\cite{Adle}:
\begin{equation}
\partial_\mu J_\mu^s = \frac{N_f g_f g_g^2}{16 \pi^2} Tr G^{\mu \nu} \tilde G_{\mu \nu}
\end{equation}
with the flavor number of fermions $N_f$ and this reflects
degenerated multiple vacua. This illustrates mass generation by
the surface effect due to the field configurations with parallel
charge electric and magnetic fields.
If $\nu = \rho_m/\rho_G$ is introduced from (\ref{tonu}) and (\ref{toma}),
a condition $\Theta \nu = 10^{-61}$ is satisfied and is consistent with the flat universe
condition $\Omega = 1 - 10^{-61}$.
$\Theta$ values parameterized by $\Theta =
10^{-61} \rho_G/\rho_m$ are consistent with the observed results,
$\Theta < 10^{-9}$ in the electric dipole moment of the neutron \cite{Alta}
and $\Theta \simeq 10^{-3}$ in the neutral kaon decay \cite{Chri}.
The condition $\Theta \nu = 10^{-61}$ is related to the instanton mechanism
represented by the tiny tunneling amplitude $e^{-S}$
with the Euclidean action $S = \Theta \nu = 10^{-61}$ in the Euclidean spacetime.

The topological winding number $\nu$ is related to the intrinsic quantum number $n_m$ by $\nu = 1/n_m^{8}$.
The intrinsic principal number $n_m$ is also connected with $N_{sp}$ and $N_{sd}$:
$n_m^2 = N_{sp}$, $N_{sp}^2 = N_{sd}$, and $N_{sp}^{4} = 1/\nu$.
The relation between the intrinsic radius and the intrinsic quantum number might be ascribed by
\begin{math}
r_i = r_{0i} / n_m^2
\end{math}
with the radius $r_{0i} = 1/m_f g_f \alpha_g \simeq N_{sp}/M_G$. Intrinsic
quantum numbers are exactly analogous to extrinsic quantum
numbers. The extrinsic principal number $n$ for the nucleon is
related to the nuclear mass number $A$ or the baryon quantum
number $B > 1$: $n^2 = A^{1/3}$, $n^4 = A^{2/3}$, $n^6 = B = A$.
Intrinsic quantum numbers introduced above are defined in the following
subsections. The relation between the nuclear radius and the
extrinsic quantum number is outlined by
\begin{equation}
r = r_0 A^{1/3} = r_0 n^2
\end{equation}
with the radius $r_0 = 1.2$ fm and the nuclear principal number
$n$. This is analogous to the atomic radius $r_e = r_0 n_e^2$ with
the atomic radius $r_0$ and the electric principal number $n_e$:
the atomic radius $r_0 = 1/2 m_e \alpha_y$ is almost the same with
the Bohr radius $a_B = 1/m_e \alpha_e = 0.5 \times 10^{-8}$ cm.
These concepts are related to the constant nuclear density $n_B =
3/4 \pi r_0^3 = 1.95 \times 10^{38} \ \textup{cm}^{-3}$ or
Avogadro's number $N_A = 6.02 \times 10^{23} \ \textup{mol}^{-1}$
and to the constant electron density $n_e = 3/4 \pi r_e^3 = 6.02
\times 10^{23} Z \rho_m/A$ with the matter energy density $\rho_m$ in the unit of
$\textup{g/cm}^3$ where the possible relation is $r_e = r_0
L^{1/3} = r_0 n_e^2$ with the lepton number $L$.

$\Theta$ values according to (\ref{thvu}) become
$\Theta_{Pl} \approx 10^{61}$, $\Theta_{EW} \approx 10^{-4}$, $\Theta_{QCD}
\approx 10^{-12}$,
and $\Theta_{0} \approx 10^{-61}$ at different stages.
The scope of $\Theta = 10^{61} \sim 10^{-61}$ corresponds to the scope of $\nu = 10^{-122} \sim 10^{0}$
to satisfy the flat universe condition $\nu \Theta = 10^{-61}$:
the maximum quantization number $N_{sp} \simeq N_R \simeq 10^{30}$ and $N_G \simeq 4 \pi N_R^3/3 \simeq 10^{91}$.
The maximum wavevector mode $N_R = (\rho_G/\Theta \rho_m)^{1/2} = 10^{30}$ of the gravitational vacuum is obtained.
These describe possible dualities between intrinsic quantum numbers and extrinsic quantum numbers:
$n_m$ and $n$, $N_{sp}^3$ and $A$, and $1/\nu$ and $A^{4/3}$ for baryons.

Fermion mass generation from the vacuum is described by $\rho_m
\simeq \rho_f \simeq \rho_c \simeq 10^{-61} \rho_G/\Theta$ with
the W boson mass density $\rho_G = M_W^4 \approx 10^8 \
\textup{GeV}^4$ at the weak scale and baryon mass generation by
$\rho_B \equiv \Omega_B \rho_c \simeq 10^{-61} \Omega_B \rho_G/\Theta$ with the
gluon mass density $\rho_G \approx 10^{-2} \ \textup{GeV}^4$ at
the strong scale. $\Theta$ terms as the surface terms modify the
original GWS model \cite{Glas} for weak interactions and the
original QCD for strong interactions \cite{Frit}, which have the
problem in the fermion mass violating gauge invariance, and
suggest mass generation as the nonperturbative breaking of
gauge and chiral invariance through DSSB.

\subsection{Fundamental Constants, Cosmological Parameters, and Conservation Laws}

One of the major developments in modern physics is the understanding of fundamental
constants and conservation laws for fundamental forces governing the universe.
Although many of them are clarified, there still exist some mysterious fundamental
constants and cosmological parameters in nature, which make profound underlying
principles of the universe be complicated. Some of them may not be really fundamental
if the origins of them can be traced. Since quantum gauge theories for fundamental
forces hold commonly absolute underlying principles such as special relativity,
quantum mechanics, and gauge invariance, fundamental constants such as the Planck
constant ($h$) and the light velocity ($c$) originate from underlying principles,
quantum mechanics and special relativity. In this context, the other fundamental
constants and cosmological constants encountered in physics are considered in depth
with relations to conservation laws associated with fundamental forces as the
consequence of gauge invariance. Various quantum numbers are quantitatively discussed
from the viewpoint of this approach under the constraint of the flat universe $\Omega
-1 = - 10^{-61}$: intrinsic quantum numbers such as color, isospin, and spin; conserved
quantum numbers such as the number of gauge bosons, the number of baryons, and the
number of photons; fundamental and cosmological constants such as Hubble's constant,
Avogadro's number, and coupling constants. Conservation laws for fundamental forces,
possible duality between intrinsic and extrinsic spacetime, and the relation between
time and gauge boson are addressed.

\subsubsection{Intrinsic Quantum Numbers}

The difference number $N_{sd}$ in intrinsic two-space dimensions suggests the
introduction of a degenerated particle number $N_{sp}$ in the intrinsic radial
coordinate and an intrinsic principal number $n_m$; particle quantum numbers are connected
by the relation $n_m^4 = N_{sp}^2 = N_{sd}$ and the Dirac quantization condition
\begin{equation}
\sqrt{g_f} g_g g_{gm} = 2 \pi N_{sp}
\end{equation}
is satisfied. The $N_{sp}$ is thus the degenerated
state number in the intrinsic radial coordinate that has the same principal number
$n_m$. The intrinsic principal quantum number $n_m$ consists of three quantum numbers,
that is, $n_m = (n_c, n_i, n_s)$ where $n_c$ is the intrinsic principal quantum number
for the color space, $n_i$ is the intrinsic principal quantum number for the isospin
space, $n_s$ is the intrinsic principal quantum number for the spin space. Intrinsic
quantum numbers $(n_c, n_i, n_s)$ take integer numbers. A fermion therefore possesses
a set of intrinsic quantum numbers $(n_c, n_i, n_s)$ to represent its intrinsic
quantum states.

The concept automatically adopts the three types of intrinsic angular momentum
operators, $\hat C$, $\hat I$, and $\hat S$, when intrinsic potentials for color,
isospin, and spin charges are central so that they depend on the intrinsic radial
distance: for instance, the color potential in strong interactions is dependent on the
radial distance. The intrinsic spin operator $\hat S$ has a magnitude square $\langle
S^2 \rangle = s (s + 1)$ and $s = 0, 1/2, 1, 3/2 \cdot \cdot \cdot (n_s-1)$. The third
component of $\hat S$, $\hat S_z$, has half integer or integer quantum number in the
range of $- s \sim s$ with the degeneracy $2s + 1$. The intrinsic isospin operator
$\hat I$ analogously has a magnitude square $\langle I^2 \rangle = i (i + 1)$ and $i =
0, 1/2, 1, 3/2 \cdot \cdot \cdot (n_i-1)$. The third component of $\hat I$, $\hat
I_z$, has half integer or integer quantum number in the range of $- i \sim i$ with the
degeneracy $2i + 1$. The intrinsic color operator $\hat C$ analogously has a magnitude
square $\langle C^2 \rangle = c (c + 1)$ and $c = 0, 1/2, 1, 3/2 \cdot \cdot \cdot
(n_c-1)$. The third component of $\hat C$, $\hat C_z$, has half integer or integer
quantum number in the range of $- c \sim c$ with the degeneracy $2c + 1$. The
principal number $n_m$ in intrinsic space quantization is very much analogous to the
principal number $n$ in extrinsic space quantization and the intrinsic angular momenta
are analogous to the extrinsic angular momentum so that the total angular momentum has
the form of
\begin{equation}
\vec J = \vec L + \vec S + \vec I + \vec C  ,
\end{equation}
which is the extension of
the conventional total angular momentum $\vec J = \vec L + \vec S$. The intrinsic
principal number $n_m$ denotes the intrinsic spatial dimension or radial quantization:
$n_c = 3$ represents strong interactions as an $SU(3)_C$ gauge theory, $n_i = 3$
represents weak interactions as an $SU(3)_I$ gauge theory, $n_s = 2$ represents
possible spin interactions as an $SU(2)_S$ gauge theory. For QWD as the $SU(3)_I$
gauge theory, there are nine weak gauge bosons ($n_i^2 = 3^2 = 9$), which consist of
one singlet gauge boson $A_0$ with $i=0$, three degenerate gauge bosons $A_1 \sim A_3$
with $i=1$, and five degenerate gauge bosons $A_4 \sim A_8$ with $i=2$; for the GWS
model as the $SU(2)_L \times U(1)_Y$ gauge theory, one singlet gauge boson $A_0$ with
$i=0$, three gauge bosons $A_1 \sim A_3$ with $i=1$, and one gauge boson $A_8$ with
$i=2$ are required. For QCD as the $SU(3)_C$ gauge theory, there are nine gluons
($n_c^2 = 3^2 = 9$), which consist of one singlet gluon $A_0$ with $c=0$, three
degenerate gluons $A_1 \sim A_3$ with $c=1$, and five degenerate gluons $A_4 \sim A_8$
with $c=2$; for QND as the $SU(2)_N \times U(1)_Z$ gauge theory, one singlet gluon
$A_0$ with $c=0$, three gluons $A_1 \sim A_3$ with $c=1$, and one gauge boson $A_8$
with $c=2$ are required. Similar scheme might be applicable to spin interactions,
which will be further discussed in the subject of quantum spindynamics as a gauge
theory. One explicit evidence of colorspin and isospin angular momenta is strong
isospin symmetry in nucleons, which is postulated as the combination symmetry of
colorspin and weak isospin in this scheme. Another evidence is the nuclear magnetic
dipole moment: the Lande spin g-factors of the proton and neutron are respectively
$g_s^p = 5.59$ and $g_s^n = - 3.83$, which are shifted from $2$ and $0$, because of
contributions from color and isospin degrees of freedom as well as spin degrees of
freedom. The mass ratio of the proton and the constituent quark, $m_p/m_q \sim 2.79$,
thus represents three intrinsic degrees of freedom of color, isospin, and spin. In
fact, the extrinsic angular momentum associated with the intrinsic angular momentum
may be decomposed by $\vec L = \vec L_i + \vec L_c + \vec L_s$ where $\vec L_i$ is the
angular momentum originated from the isospin charge, $\vec L_c$ is the angular
momentum originated from the color charge, and $\vec L_s$ is the angular momentum
originated from the spin charge. This is supported by the fact that the orbital
angular momentum $l_c$ of the nucleon has the different origin from the color charge
with the orbital angular momentum $l_i$ of the electron from the isospin charge since
two angular momenta have opposite directions from the information of spin-orbit
couplings in nucleus and atoms. Extrinsic angular momenta have extrinsic parity
$(-1)^l = (-1)^{(l_c + l_i + l_s)}$, intrinsic angular momenta have intrinsic parity
$(-1)^{(c + i + s)}$, and the total parity becomes $(-1)^{(l + c + i + s)}$ for
electric moments while extrinsic angular momenta have extrinsic parity $(-1)^{(l + 1)}
= (-1)^{(l_c + l_i + l_s + 1)}$, intrinsic angular momenta have intrinsic parity
$(-1)^{(c + i + s + 1)}$, and the total parity becomes $(-1)^{(l + c + i + s + 1)}$
for magnetic moments.

Fermions increase their masses by decreasing their intrinsic
principal quantum numbers from the higher ones at higher energies to the lower ones at
lower energies. The coupling constant $\alpha_g$ of a non-Abelian gauge theory is
strong for the small $N_{sd}$ and is weak for the large $N_{sd}$ according to the
renormalization group analysis. The vacuum energy is described by the zero-point
energy in the unit of $\omega/2$ with the maximum number $N_{sd} \simeq 10^{61}$ and
the vacuum is filled with fermion pairs of up and down colorspins, isospins, or spins,
whose pairs behave like bosons quantized by the unit of $\omega$: this is analogous to
the superconducting state of fermion pairs. The intrinsic particle number $N_{sp}
\simeq 10^{30}$ (or $B \simeq 10^{-12}$, $L \simeq 10^{-9}$) characterizes
gravitational interactions for fermions with the mass $10^{-12}$ GeV, $N_{sp} \simeq
10^{6}$ (or $L_e \simeq 1$) characterizes weak interactions for electrons, and $N_{sp}
\simeq 1$ (or $B \simeq 1$) characterizes strong interactions for nucleons.
Fundamental particles known as leptons and quarks are hence postulated as composite
particles with the color, isospin, and spin quantum numbers; the quark is a color
triplet state but the lepton is a color singlet. Note that if $N_{sp} > 1$ (or $B <
1$), it represents a pointlike fermion and if $N_{sp} < 1$ (or $B > 1$), it represents
a composite fermion.

\subsubsection{Extrinsic Quantum Numbers}

In this scheme, vacuum and matter energies are spatially quantized as well as photon
and phonon energies. The vacuum represented by massive gauge bosons is quantized
by the maximum wavevector mode $N_R = i/(\Omega -1)^{1/2} \approx 10^{30}$ and the
total gauge boson number $N_G = 4 \pi N_R^3/3 \approx 10^{91}$; the wavevector $k_G =
(E^2 - M_G^2)^{1/2}$ is quantized by the maximum wavevector mode $N_R$ if $M_G > E$.
The maximum wavevector mode $N_R \approx 10^{30}$ is manifest since the universe size
is $R_{Pl} \approx 10^{-3}$ cm at the Planck scale $l_{Pl} \approx 10^{-33}$ cm and
the universe size is $R_{0} \approx 10^{28}$ cm at the present scale $l_{Pl} \approx
10^{-3}$ cm if the universe is extremely flat, $\Omega - 1 = - 10^{-61}$. Baryon
matter represented by massive baryons is quantized by the maximum wavevector mode
$N_F \approx 10^{26}$ and the total baryon number $B = N_B = 4 \pi N_F^3/3 \approx
10^{78}$; the wavevector $k_B = (E^2 - m_B^2)^{1/2}$ is quantized by the wavevector
mode $N_F$ if the baryon mass $m_B$ is bigger than its energy $E$. Baryon matter
quantization is consistent with the nuclear matter number density $n_n \approx n_B
\approx 1.95 \times 10^{38} \ \textup{cm}^{-3}$ and Avogadro's number $N_A = 6.02
\times 10^{23} \ \textup{mol}^{-1} \approx 10^{19} \ \textup{cm}^{-3}$ in the matter;
the baryon number density at the nuclear interaction scale $10^{-1}$ GeV is $10^{26} \
\textup{cm}^{-3}$ in the universe size $R_{QCD} \approx 10^{17}$ cm, whose volume
$10^{51} \ \textup{cm}^{3}$ is $10^{12}$ times bigger than the matter volume $10^{39}
\ \textup{cm}^{3}$. Electrons with the mass $0.5$ MeV might be similarly quantized by
$N_F \approx 10^{27}$ and the total number $10^{81}$ if the electron number is
conserved under the assumption of $\Omega_e = \rho_e/\rho_c \approx 1$. The maximum
wavevector mode $N_F$ is close to $10^{30}$ if the mass quantization unit of fermions
$10^{-12}$ GeV is used rather than the mass unit of baryons $0.94$ GeV under the assumption
of the fermion number conservation $N_f \simeq 10^{91}$: the fermion with the mass
$10^{-12}$ GeV has much bigger intrinsic quantum number $N_{sd}$ than the baryon with
the mass $0.94$ GeV has. Massless gauge bosons (photons) are quantized by the maximum
wavevector mode $N_\gamma \approx 10^{29}$ and the total massless boson number $N_{t
\gamma} = 4 \pi N_\gamma^3/3 \approx 10^{88}$. CMBR is the conclusive evidence for
massless gauge bosons with the number $N_{t \gamma} \approx 10^{88}$. Massless phonons
in the matter space are quantized by the maximum wavevector mode (Debye mode) $N_D \approx 10^{25}$ and
the total phonon number $N_{t p} = 4 \pi N_D^3/3 \approx 10^{75}$. These total
particle numbers $N_G \approx 10^{91}, N_B \approx 10^{78}$, $N_{t \gamma} \approx
10^{88}$, and $N_{t p} \approx 10^{75}$  are conserved good quantum numbers. Vacuum,
matter, photon, and phonon energies are also thermodynamically quantized. Quantum
states of vacuum, matter, photons, and phonons have average occupation numbers $f_b =
1 /(e^{(E - \mu)/T} - 1)$ for gauge bosons, $f_f = 1 /(e^{(E - \mu)/T} + 1)$ for
baryons, $f_\gamma = 1 /(e^{E/T} - 1)$ for photons, and $f_p = 1 /(e^{E/T} - 1)$ for
phonons under the assumption of free particles in thermal equilibrium.

\subsubsection{Fundamental Constants and Cosmological Parameters}

Coupling constants for weak and strong interactions are unified around $10^{2}$ GeV or slightly higher energy (rather than the order of $10^{15}$ GeV):
$\alpha_h = \alpha_i = \alpha_s \simeq 0.12$ \cite{Roh2,Roh3}.
The unification at the order of a TeV energy is consistent with recent GUT \cite{Poma}.
In terms of the Weinberg weak mixing angle $\sin^2 \theta_W = 1/4$ and the strong mixing angle $\sin^2 \theta_R = 1/4$,
coupling constant hierarchies for weak and strong interactions are respectively obtained.
Electroweak coupling constants are $\alpha_z = i_f^z \alpha_i = \alpha_i/3 \simeq 0.04$,
$\alpha_w = i_f^w \alpha_i = \alpha_i/4 \simeq 0.03$, $\alpha_y = i_f^y \alpha_i = \alpha_i/12 \simeq 0.01$, and
$\alpha_e = i_f^e \alpha_i = \alpha_i/16 \simeq 1/133$ as symmetric isospin interactions at the weak scale
and $-2 \alpha_i/3$, $- \alpha_i/2$, $- \alpha_i/6$, and $- \alpha_i/8$ as asymmetric isospin interactions:
$i_f^w = \sin^2 \theta_W$ and $i_f^e = \sin^4 \theta_W$.
Strong coupling constants for baryons are
$\alpha_b = c_f^b \alpha_s = \alpha_s/3 $, $\alpha_n = c_f^n \alpha_s = \alpha_s/4$,
$\alpha_z = c_f^z \alpha_s = \alpha_s/12$, and $\alpha_f = c_f^f \alpha_s = \alpha_s/16$
as symmetric color interactions and $- 2 \alpha_s/3 $, $- \alpha_s/2 $, $- \alpha_s/6$,
and $- \alpha_s/8$ as asymmetric color interactions:
$c_f^n = \sin^2 \theta_R$ and $c_f^f = \sin^4 \theta_R$.
The charge (isospin or color) factors introduced are
$i^s_f = (i_f^z, i_f^w, i_f^y, i_f^e) = c^s_f = (c_f^b, c_f^n, c_f^z, c_f^f) = (1/3, 1/4, 1/12, 1/16)$
for symmetric repulsive interactions and $i^a_f = c^a_f = (-2/3, -1/2, -1/6, -1/8)$ for asymmetric attractive interactions.
The symmetric charge factors reflect intrinsic even parity with repulsive force while
the asymmetric charge factors reflect intrinsic odd parity with attractive force;
this suggests electromagnetic duality.
Asymmetric configuration for attractive force is confined inside particle while symmetric configuration for repulsive force is
appeared on scattering or decay processes.
This may hint the underlying principles of nature known as the duality both between the intrinsic and extrinsic space
and duality between electricity and magnetism.
The coupling constant chain is $\alpha_g \rightarrow \alpha_h \rightarrow \alpha_i \rightarrow \alpha_s$ for gravitation, grand unification, weak, and strong interactions respectively.

The baryon matter density $\rho_B$ may be connected with a baryon coupling constant $\alpha_z$ by
\begin{equation}
\label{made}
\rho_B = A m_n/V = A m_n/(4 \pi r_0^3 A/3) = 3 m_n^4 \alpha_z^3 / \pi
\end{equation}
if the nucleon number $A = B$, the nucleus radius $r = r_0 A^{1/3}$, and the mass radius $r_0 = 1/2 m_n \alpha_z = 1.2$ fm are used.
In the estimation of the mass radius $r_0$, the factor $2$ reflects the color degeneracy number of nucleons and
the nuclear number density $n_B = A/V = B/V = 3/(4 \pi r_0^3) \approx 1.95 \times 10^{38} \ \textup{cm}^{-3}$ reflects the baryon number conservation
as the result of the $U(1)_Z$ gauge theory.
(\ref{made}) is thus useful to estimate to the size, total mass, mass density, and coupling constant of baryon matter
if some parts of them are known.

Table \ref{fuco} summarizes fundamental and cosmological constants
in quantum cosmology: the gauge boson mass $M_G$, the effective
coupling constant $G_G \simeq \sqrt{2} g^2/8 M_G^2$, the gauge
boson number density $n_G \simeq M_G^3$, the vacuum energy density
$V_e (\bar \phi) \simeq M_G^4$, the cosmological constant
$\Lambda_e \simeq 8 \pi G_N M_G^4$, the Hubble constant $H_e =
(\Lambda_e/3)^{1/2} \simeq (8 \pi G_N M_G^4/3)^{1/2}$, the baryon
number density $n_B$, the baryon mass density $\rho_B$, the
electron number density $n_e$, the electron mass density $\rho_e$,
the photon energy $E_\gamma$, the photon number density $n_{t
\gamma} \simeq 2 \zeta(3) T^3/\pi^2$, the photon energy density
$\epsilon_\gamma$, the phonon number density $n_{t p}$, the phonon
energy density $\epsilon_p$, the $\Theta$ constant $\Theta \simeq
10^{-61} \rho_G/\rho_m$, and the topological constant $\nu \simeq
\rho_m/\rho_G$.  The values of the baryon (electron) number
density and mass density represent ones when the vacuum volume is
used while the values within parentheses represent ones when only
the baryon (electron) matter volume is used.

\subsubsection{Conservation Laws}

Total particle numbers such as the gauge boson number $N_G \approx 10^{91}$, the
baryon number $N_B \approx 10^{78}$, the electron number $N_e \approx 10^{81}$, the
photon number $N_{t \gamma} \approx 10^{88}$, and the phonon number $N_{t p} \approx
10^{75}$ are conserved good quantum numbers as described above. The matter current at
the Planck scale, the (V - A) current and the electromagnetic current at the
electroweak scale, the baryon current and the proton current at the strong scale, and
the lepton current at the present scale might be conserved currents at different
energy scales. The proton number conservation is the consequence of the $U(1)_f$ gauge
theory just as the electron number conservation is the consequence of the $U(1)_e$
gauge theory. In gravitational interactions, the predicted typical lifetime for a
particle with the mass $1$ GeV is $\tau_p = 1/\Gamma_p \simeq 1/G_N^2 m^5 \approx
10^{50}$ years using the analogy of the lifetime of the muon $\tau_\mu = 192
\pi^3/G_F^2 m_\mu^5$ in weak interactions. Therefore, in the proton decay $p
\rightarrow \pi^0 + e^+$ at the energy $E << M_{Pl}$, the proton would have much
longer lifetime than $10^{32}$ years if the decay process is gravitational. In fact,
the lower bound for the proton lifetime is $10^{32}$ years at the moment. If the
electric charge is completely conserved, the electron can not decay. The present lower
bounds for the electron lifetime are bigger than $10^{21}$ years for the electron
decay into neutral particles and $10^{25}$ years for the decay $e^{-} \rightarrow
\gamma + \nu_e$ \cite{Stei}. The baryon number conservation is the result of the
$U(1)_Z$ gauge theory for strong interactions just as the lepton number conservation
is the result of the $U(1)_Y$ gauge theory for weak interactions. The bound of lepton
number nonconservation process is expressed by the branching ratio $B (K^+ \rightarrow
\pi^- e^+e^+) < 10^{-8}$ or $B (\mu^-N \rightarrow e^+N') < 7 \times 10^{-11}$ and the
bound of lepton flavor violation is shown by $B (\mu^- \rightarrow e^- \gamma) < 5
\times 10^{-11}$, $B (\mu^- \rightarrow e^- \gamma \gamma) < 7 \times 10^{-11}$, or $B
(\mu^- \rightarrow 3 e^-) < 10^{-13}$. Conservation laws of the baryon number, lepton
number, and electric charge number are good in weak, strong, and present interactions
but they would be nonperturbatively violated in gravitational interactions: there is
possibility for such violation even at much lower energy although they are extremely
small. Discrete symmetries such as parity (P), charge conjugate (C), time reversal
(T), and charge conjugate and parity (CP) are conserved perturbatively but are
violated nonperturbatively during DSSB. The violation is analogous to the
nonconservation of the (V + A) current in weak interactions and to the axial vector
current in strong interactions. The breaking of discrete symmetries through the
condensation of singlet gravitons might cause the matter-antimatter asymmetry at the
Planck scale: the parameter $\Theta_{Pl} \simeq 10^{61}$. The breaking of discrete
symmetries through the condensation of intermediate vector bosons causes the
lepton-antilepton asymmetry at the weak scale: the parameter $\Theta_{EW} \simeq
10^{-4}$. The absence of the right-handed neutrino shows P violation \cite{Lee} and
the decay of the neutral kaon shows CP violation \cite{Chri}.

The breaking of discrete symmetries through the condensation of
singlet gluons causes the baryon-antibaryon asymmetry at the
strong scale: the parameter $\Theta_{QCD} \simeq 10^{-12}$. Hadron
mass spectra support this scheme since pseudoscalar and vector
mesons are observable while their parity partners, scalar and
pseudovector mesons, are not observable; similarly, there are no
baryon octet and decuplet parity partners. This, of course,
resolves the $U(1)_A$ problem; the absence of the $U(1)_A$
particle is due to the nonconservation of the color axial vector
current. Even in electromagnetic force, discrete symmetries might
be slightly violated. There is, for instance, no explicit
electric-magnetic symmetries in Maxwell's equations: neither
magnetic monopole nor scalar potential for the magnetic field is
observed. The electric dipole moment and the electron mass are
also good examples for the breaking of discrete symmetries. A
photon might have mass although it is extremely small and the
present upper bound for the photon mass is $m_\gamma < 10^{-16}$
GeV \cite{Davi}. If baryon and electron numbers $N_B \simeq N_e
\simeq 10^{78}$ are simultaneously conserved, $\rho_B \simeq 10^3
\ \rho_e$, which contradicts with $\rho_B \approx \rho_e \approx
\rho_c$ at the Planck scale. This implies the nonconservation of
baryon and lepton numbers at the Planck scale. The conservation of
the fermion number $N_f \simeq 10^{91}$ in the unit of mass
$10^{-12}$ GeV is good at the Planck scale under the assumption
with no supersymmetry and higher dimensions; quarks and leptons
are postulated as composite particles composed  from more
fundamental particles. The baryon number is not conserved for
color axial singlet hadrons just as the lepton number is not
conserved for right-handed isospin singlet leptons. Lorentz
invariance or TCP invariance seems to be related to the flat
universe condition $\Omega - 1 = - 10^{-61}$, which might cause
the possible, tiny violation of Lorentz invariance or TCP
invariance in the order of $10^{-30}$ due to DSSB; the TCP theorem
is at the moment supported by the mass difference $(K^0 - \bar
K^0)$ which is less than $6 \times 10^{-19}$ as a fraction of
$K^0$ mass, the mass difference $(\pi^+ - \pi^-) \approx 1.7
\times 10^{-3}$ MeV, and the lifetime equalities for the muon,
pion, kaon and their antiparticles respectively \cite{Meye}. Table
\ref{cola} summarizes the overview of conservation laws for
fundamental forces and Table \ref{coga} shows relations between
conservation laws and gauge theories in weak and strong
interactions.

\subsubsection{Duality between Intrinsic and Extrinsic Spacetime}

Each elementary particle such as the lepton or quark has both
intrinsic and extrinsic properties. Intrinsic properties such as
spin, charge, and mass, are all represented by good quantum
numbers such as the intrinsic principal quantum number, intrinsic
angular momentum, and the third component of intrinsic angular
momentum for color, isospin, and spin: $(n_c, c, m_c)$, $(n_i, i,
m_i)$, and $(n_s, s, m_s)$ respectively. Identical particles also
possess good extrinsic quantum numbers like space quantization:
the total wave function of a fermion $\psi (\vec r, \vec C, \vec
I, \vec S) = \psi (\textup{space}) \psi (\textup{color}) \psi
(\textup{isospin}) \psi (\textup{spin})$ follows the Pauli
exclusion principle. The description above suggests duality
property between intrinsic and extrinsic spacetime before DSSB.
Intrinsic and extrinsic quantum numbers have one to one
correspondence since the intrinsic principal number and extrinsic
principal number are analogous. The total principal quantum may be
introduced by $n_t = n_m n$ with the intrinsic principal number
$n_m$ and the extrinsic quantum number $n$. The maximum total
principal quantum number $n_t$ has the order of $10^{15}$ so that
the maximum quantum numbers are $N_{sp} \approx 10^{30}$, $N_{sd}
\approx 10^{61}$, $\nu \approx 10^{122}$. Intrinsic and extrinsic
angular momenta form total angular momentum $\vec J = \vec C +
\vec I + \vec S + \vec L$. The inside potential has the distance
dependence $r^{l}$ while the outside potential has the distance
dependence $1/r^{l+1}$. The dual properties between intrinsic and
extrinsic orbital angular momenta may be identified and be
reflected by the uncertainty principles, $\Delta c_z \Delta
\varphi_c \geq 1/2$, $\Delta i_z \Delta \varphi_i \geq 1/2$, and
$\Delta s_z \Delta \varphi_s \geq 1/2$, between the longitudinal
components of intrinsic angular momenta and azimuthal angles.
There are dual properties between electricity and magnetism
\cite{Dira,Mand,Seib} before DSSB even though one part of them
disappears in order to satisfy parity during DSSB: the dual
property might exist before DSSB while the electric monopole
exists but the magnetic monopole does not exist after DSSB. The
Dirac quantization condition $\sqrt{g_f} g_g g_{gm} = 2 \pi
N_{sp}$ is related to intrinsic quantum numbers as seen in the
dual pairing mechanism of mass generation. There are even-odd dualities
between discrete symmetries P, C, T, and CP for gauge bosons as
well as fermions, which might be restored before DSSB.
Nonperturbative symmetry breaking described above might thus
suggest the restoration of perfect symmetry before DSSB if
supersymmetry between fermions and bosons, higher dimensions above
one-time and three-space dimensions, symmetry between space and time,
duality between the intrinsic-extrinsic space, duality between
discrete symmetries, and duality between electricity and magnetism
are adopted.

\subsubsection{Relation between Time and Gauge Boson Mass}

Relations among time, energy, temperature, and universe size are discussed at each
epoch in quantum cosmology.

At a radiation dominated universe of Einstein's field equation, the expansion rate is
well approximated by
\begin{math}
(\frac{\dot {R}}{R})^2 = \frac{8 \pi G_N a_R T^4}{3}
\end{math}
with the radiation density constant
$a_R = \pi^2/15 = 7.56 \times 10^{-15} \ \textup{erg} \ \textup{cm}^{-3} \ \textup{K}^{-4}$ and
the time scale is then given by
\begin{equation}
\label{timm}
t = (\frac{3}{32 \pi G_N a_R T^4})^{1/2} .
\end{equation}
Based on this scheme, however, the time scale is exactly given in terms of the gauge boson mass by
\begin{equation}
\label{temm}
t = \frac{1}{H_e} = (\frac{3}{8 \pi G_N M_G^4})^{1/2}
\end{equation}
where $M_G \sim T$. Using (\ref{temm}) and fundamental constants,
different scales of time, energy, temperature, and universe size
in the universe evolution are summarized at Table \ref{scqc}. In
this scheme, the quantization units of energy, temperature,
frequency, time, and distance in the universe are respectively
$10^{-42}$ GeV, $10^{-30}$ K, $10^{-19}$ Hz, $10^{-43}$ s, and
$10^{-33}$ cm if gravitational and present interactions are
concerned.

\section{Conclusions}

This study toward quantum gravity (QG) proposes an $SU(N)$ gauge
theory with the $\Theta$ vacuum term as a trial theory, which
suggests that a certain group $G$ for gravitational interactions
leads to a group $SU(2)_L \times U(1)_Y \times SU(3)_C$ for weak and strong
interactions through dynamical spontaneous symmetry breaking
(DSSB) leading to a current anomaly; the group chain is $G \supset
SU(2)_L \times U(1)_Y \times SU(3)_C$.
The typical predictions of QG are consistent with recent experiments,
BUMERANG-98 and MAXIMA-1:
the flat universe, inflation, vacuum energy, dark matter, repulsive force, CMBR, etc.
DSSB consists of two simultaneous
mechanisms; the first mechanism is the explicit symmetry breaking
of gauge symmetry, which is represented by the gravitational
factor $g_f$ and the gravitational coupling constant $g_g$, and
the second mechanism is the spontaneous symmetry breaking of gauge
fields, which is represented by the condensation of singlet gauge
fields. Newton gravitation constant $G_N$ originates from the
effective coupling constant for massive gravitons,
$\frac{G_N}{\sqrt{2}} = \frac{g_f g_g^2}{8 M_G^2}$ with $M_G =
M_{Pl} \approx 10^{19}$ GeV: the effective coupling constant chain
is $G_N \supset G_F \times G_R$ for gravitation, weak, and strong
interactions respectively. This scheme relates the effective
cosmological constant to the effective vacuum energy associated
with massive gauge bosons, $M_G^2 = M_{Pl}^2 - g_f g_g^2
\langle \phi \rangle^2$, and provides a plausible explanation for both the
present small and the early large value of the cosmological
constant; the condensation of the singlet gauge field $\langle \phi \rangle$
induces the current anomaly and subtracts the gauge boson mass as
the system expands. This proposal thus suggests a viable solution
toward such longstanding problems as the quantization of gravity
and the cosmological constant.

This paper demonstrates that the present universe in the mixed phase of DSSB can successfully be examined by
quantum tests predicted by gauge theory.
The dual Meissner effect in the superconducting state explains why gravitational
waves are not easily observed
by showing the exclusion of the gravitational electric field analogously to the exclusion of the magnetic field in the electric superconductivity;
the massive graviton has the Planck mass.
The universe scale $R(t) = R(0) \ \exp (\int^t_0 H_e dt)$ with the effective Hubble
constant $H_e = (\Lambda_e/3)^{1/2} = (8 \pi G_N M_G^4/3)^{1/2}$: the radius of
spatial curvature $R_c = i M_G^{-1}/(\Omega-1)^{1/2} = N_R/M_G \approx (10^{-11} \
\textup{GeV})^{-1} \approx 10^{-3}$ cm and the gauge boson mass $M_G \approx 10^{19}$
GeV at the Planck epoch $t_{Pl} \approx 10^{-43}$ s and $R_c = 1/H_0 \approx (10^{-42}
\ \textup{GeV})^{-1} \approx 10^{28}$ cm and $M_G \approx 10^{-12}$ GeV at the present
epoch $t_0 \approx 10^{17}$ s. The maximum wavevector mode of massive gauge bosons
$N_R = i/(\Omega - 1)^{1/2} \approx 10^{30}$ is here introduced so as to resolve the
problems of the size, flatness, and horizon of the universe.
The matter-antimatter asymmetry in the universe may be explained by the nonconservation of antibaryon current and
the breaking of discrete symmetries, charge conjugation (C), parity (P), charge conjugation and parity (CP), and time reversal (T) during phase transition.
The baryon asymmetry $\delta_B = \frac{N_B}{N_{t \gamma}} \approx 10^{-10}$ occurred
at the strong scale is almost kept in constant
at the later stages of lower energies, according to the present baryon asymmetry.
This is also confirmed in terms of the Avogadro's number of atoms $N_A = 6.02 \times 10^{23} \ \textup{mol}^{-1}$ and the nuclear number density
$n_n = 1,95 \times 10^{38} \ \textup{cm}^{-3}$.
The expansion of the universe suggests a gauge theory with the nearly massless gauge boson
responsible for a new type of force:
the mass $M_G \approx 10^{-12}$ GeV and the effective coupling constant
$G_S = \sqrt{2} g_r^2 / 8 M_G^2 \approx 10^{24} \ \textup{GeV}^{-2}$,
which is $10^{61}$ times stronger than $G_N$.
Non-zero mass gauge bosons with the particle number $N_G \approx 10^{91}$ become mediators of nonbaryonic dark matter
and CMBR at $2.7 \ K \approx 3 \times 10^{-13}$ GeV with the photon number $N_{t \gamma} \approx 10^{88}$ illustrates the existence of massless gauge bosons during DSSB at the present epoch.
SIMPs are suggested as observable dark matter in addition to WIMPs.
Primordial nucleosynthesis also suggests dark matter and structure formation due to dark matter further supports this scheme.

The mechanism of fermion mass generation is suggested in terms of the DSSB of gauge
symmetry and discrete symmetries known as the dual pairing mechanism of the
superconducting state: $M_G = \sqrt{\pi} m_f g_f \alpha_g \sqrt{N_{sd}}$. The
difference number of fermions $N_{sd}$ in fermion mass generation represents $N_{sd} =
N_{ss} - N_{sc}$ where $N_{ss}$ is the number of singlet fermions and $N_{sc}$ is the
condensed number of paired fermions: there exists an electric-magnetic duality before
DSSB, which is closely related to quantum numbers $N_{sd}$ and $N_{sc}$. At the phase
transition, $N_{sc}$ becomes zero so that $N_{sd}$ becomes the maximum. Using
relations $M_G = \sqrt{\pi} m_f g_f \alpha_g \sqrt{N_{sd}}$ and $M^2_G = M^2_{Pl} -
g_f g_g^2 \langle \phi \rangle^2$, the zero point energy $M_{Pl}^2 = \pi m_f^2 g_f^2
\alpha_g^2 N_{ss}$ and the reduction of the zero-point energy $\langle \phi \rangle^2
= m_f^2 g_f \alpha_g N_{sc}/4$ are obtained. Since the $\Theta$ constant is
parameterized by $\Theta = 10^{-61} \ \rho_G/\rho_m$ with the vacuum energy density
$\rho_G = M_G^4$ and the matter energy density $\rho_m \simeq \rho_c \simeq 10^{-47} \
\textup{GeV}^4$, the relation between the $\Theta$ constant and the difference number
$N_{sd}$ is given by $\Theta = \pi^2 g_f^4 \alpha_g^4 m_f^4 N_{sd}^2/10^{61} \rho_c$.
In the dual pairing mechanism, gravitational electric monopole, gravitational magnetic
dipole, and gravitational electric quadrupole remain in the matter space but
gravitational magnetic monopole, gravitational electric dipole, and gravitational
magnetic quadrupole condense in the vacuum space. Antimatter particles condense in the
vacuum space while matter particles remain in the matter space as the consequence of
charge conjugation symmetry breaking: the matter-antimatter asymmetry.

The difference number of even-odd parity singlet fermions $N_{sd}$ in intrinsic
two-space dimensions suggests the introduction of a degenerated particle number $N_{sp}$
in the intrinsic radial coordinate and an intrinsic principal number $n_m$; particle
numbers are connected with the relation $n_m^4 = N_{sp}^2 = N_{sd}$ and the Dirac quantization condition
$\sqrt{g_f} g_g g_{gm} = 2 \pi N_{sp}$ is satisfied. The $N_{sp}$ is
thus the degenerated state number in the intrinsic radial coordinate that has the same
principal number $n_m$. The intrinsic principal quantum number $n_m$ consists of three
quantum numbers, that is, $n_m = (n_c, n_i, n_s)$ where $n_c$ is the intrinsic
principal quantum number for the color space, $n_i$ is the intrinsic principal quantum
number for the isospin space, $n_s$ is the intrinsic principal quantum number for the
spin space. Intrinsic quantum numbers $(n_c, n_i, n_s)$ take integer numbers. A
fermion therefore possesses a set of intrinsic quantum numbers $(n_c, n_i, n_s)$ to
represent its intrinsic quantum states. The concept automatically adopts the three
types of intrinsic angular momentum operators, $\hat C$, $\hat I$, and $\hat S$, when
intrinsic potentials for color, isospin, and spin charges are central so that they
depend on the intrinsic radial distance: for instance, the color potential in strong
interactions is dependent on the radial distance. The principal number $n_m$ in
intrinsic space quantization is very much analogous to the principal number $n$ in
extrinsic space quantization and the intrinsic angular momenta are analogous to the
extrinsic angular momentum so that the total angular momentum has the form of $\vec J
= \vec L + \vec S + \vec I + \vec C$, which is the extension of the conventional total
angular momentum $\vec J = \vec L + \vec S$. The intrinsic principal number $n_m$
denotes the intrinsic spatial dimension or radial quantization: $n_c = 3$ represents strong interactions as
an $SU(3)_C$ gauge theory, $n_i = 2$ represents weak interactions as an $SU(2)_L \times U(1)_Y$
gauge theory, $n_s = 2$ represents possible spin interactions as an $SU(2)_S$ gauge
theory. One explicit evidence of colorspin and isospin angular momenta is strong
isospin symmetry in nucleons, which is postulated as the combination symmetry of
colorspin and weak isospin in this scheme. Another evidence is the nuclear magnetic
dipole moment: the Lande spin g-factors of the proton and neutron are respectively
$g_s^p = 5.59$ and $g_s^n = - 3.83$, which are shifted from $2$ and $0$, because of
contributions from color and isospin degrees of freedom as well as spin degrees of
freedom. The mass ratio of the proton and the constituent quark, $m_p/m_q \sim 2.79$,
thus represents three intrinsic degrees of freedom of color, isospin, and spin. In
fact, the extrinsic angular momentum may be decomposed by $\vec L = \vec L_i + \vec
L_c + \vec L_s$ where $\vec L_i$ is the angular momentum originated from the isospin
charge, $\vec L_c$ is the angular momentum originated from the color charge, and $\vec
L_s$ is the angular momentum originated from the spin charge. Fermions increase their
masses by decreasing their intrinsic principal quantum numbers from the higher ones at
higher energies to the lower ones at lower energies. The coupling constant $\alpha_g$
is strong for the small $N_{sd}$ and is weak for the large $N_{sd}$ according to the
renormalization group analysis. The vacuum energy is described by the zero-point
energy in the unit of $\omega/2$ with the maximum number $N_{sd} \simeq 10^{61}$ and
the vacuum is filled with fermion pairs of up and down colorspins, isospins, or spins,
whose pairs behave like bosons quantized by the unit of $\omega$: this is analogous to
the superconducting state of fermion pairs. The intrinsic particle number $N_{sp}
\simeq 10^{30}$ (or $B \simeq 10^{-12}$, $L \simeq 10^{-9}$) characterizes
gravitational interactions for fermions with the mass $10^{-12}$ GeV, $N_{sp} \simeq
10^{6}$ (or $L_e \simeq 1$) characterizes weak interactions for electrons, and $N_{sp}
\simeq 1$ (or $B \simeq 1$) characterizes strong interactions for nucleons.
Fundamental particles known as leptons and quarks are hence postulated as composite
particles with the color, isospin, and spin quantum numbers; the quark is a color
triplet state but the lepton is a color singlet. Note that if $N_{sp} > 1$ (or $B <
1$), it represents a pointlike fermion and if $N_{sp} < 1$ (or $B
> 1$), it represents a composite fermion.

The invariance of gauge transformation provides
$\psi [\hat O_\nu] = e^{i \nu \Theta} \psi [\hat O]$ for the fermion wave function $\psi$ with the transformation of an operator $\hat O$ by
the class $\nu$ gauge transformation, $\hat O_\nu$:
the vacuum state characterized by the constant $\Theta$ is called the $\Theta$ vacuum.
The true vacuum is the superposition of all the $|\nu \rangle$ vacua with the phase $e^{i \nu \Theta}$:
$|\Theta \rangle = \sum_\nu e^{i \nu \Theta} |\nu \rangle$.
The topological winding number $\nu$ or the topological charge $q_s$ is defined by
\begin{math}
\nu = \nu_+ - \nu_- = \int \frac{g_f g_g^2}{16 \pi^2} Tr G^{\mu \nu} \tilde G_{\mu \nu} d^4 x
\end{math}
where the subscripts $+$ and $-$ denote moving particles with opposite characteristics
respectively in the presence of the gauge fields. The matter energy density generated by the
surface effect is postulated by
\begin{math}
\rho_m \simeq \rho_c \simeq \frac{g_f g_g^2}{16 \pi^2} Tr G^{\mu \nu} \tilde G_{\mu \nu}
\end{math}
which implies that the fermion mass is generated by the difference of fermion numbers
moving to backward and forward directions. In this aspect, the difference number
$N_{sd}$, the singlet fermion number $N_{ss}$, and the condensed singlet fermion
number $N_{sc}$ in intrinsic two-space dimensions respectively correspond to $\nu$,
$\nu_+$, and $\nu_-$ in three-space and one-time dimensions.  In the presence of the $\Theta$
term, the odd singlet current is not conserved due to an anomaly:
\begin{math}
\partial_\mu J_\mu^s = \frac{N_f g_f g_g^2}{16 \pi^2} Tr G^{\mu \nu} \tilde G_{\mu \nu}
\end{math}
with the flavor number of fermions $N_f$ and this reflects
degenerated multiple vacua. This illustrates mass generation by
the surface effect due to the field configurations with parallel
electric and magnetic fields. $\Theta$ values defined by $\Theta =
10^{-61} \rho_G/\rho_m$ are consistent with the observed results,
$\Theta < 10^{-9}$ in the electric dipole moment of the neutron
and $\Theta \simeq 10^{-3}$ in the neutral kaon decay.
The topological winding number $\nu = 10^{-61}/\Theta = \rho_m/\rho_G$ is related to the intrinsic
quantum number $n_m$ by $\nu = n_m^{-8}$. The intrinsic
principal number $n_m$ is also connected with $N_{sp}$ and
$N_{sd}$: $n_m^2 = N_{sp}$, $N_{sp}^2 = N_{sd}$, and $N_{sp}^{4} =
1/\nu$. The relation between the intrinsic radius and the
intrinsic quantum number might be ascribed by $r_i = r_{0i} /
n_m^2$ with the radius $r_{0i} = 1/m_f g_f \alpha_g \simeq N_{sp}/M_G$.
Intrinsic quantum numbers are exactly analogous to extrinsic
quantum numbers. The extrinsic principal number $n$ for the
nucleon is related to the nuclear mass number $A$ or the baryon
quantum number $B > 1$: $n^2 = A^{1/3}$, $n^4 = A^{2/3}$, $n^6 = B
= A$. The relation between the nuclear radius and the extrinsic
quantum number is outlined by $r = r_0 A^{1/3} = r_0 n^2$ with the
radius $r_0 = 1.2$ fm and the nuclear principal number $n$ and is
analogous to the atomic radius $r_e = a_0 n_e^2$ with the atomic
radius $a_0 = 1/2 m_e \alpha_y$ or the Bohr radius $a_B = 1/m_e
\alpha_e = 0.5 \times 10^{-8}$ cm and the electric principal
number $n_e$. These concepts are related to the constant nuclear
density $n_B = 1.95 \times 10^{38} \ \textup{cm}^{-3}$ and
Avogadro's number $N_A = 6.02 \times 10^{23} \ \textup{mol}^{-1}$.
$\Theta$ values indicating DSSB become $\Theta_{Pl} \approx
10^{61}$, $\Theta_{EW} \approx 10^{-4}$, $\Theta_{QCD} \approx
10^{-12}$, and $\Theta_{0} \approx 10^{-61}$ at different stages.
$\Theta = 10^{61} \sim 10^{-61}$ corresponds to $\nu = 10^{-122}
\sim 10^{0}$ to satisfy the flat universe condition $\nu \Theta =
10^{-61}$: the maximum quantization number $N_{sp} \simeq N_R \simeq
10^{30}$ and $N_G \simeq 4 \pi N_R^3/3 \simeq 10^{91}$. The maximum
wavevector mode $N_R = (\rho_G/\Theta \rho_B)^{1/2} = 10^{30}$ of
the gravitational vacuum is obtained. These describe possible
dualities between intrinsic quantum numbers and extrinsic quantum
numbers: $n_m$ and $n$, $N_{sp}^3$ and $A$, and $1/\nu$
and $A^{4/3}$ for baryons. Fermion mass generation from the vacuum
is described by $\rho_m \simeq \rho_f \simeq \rho_c \simeq
10^{-61} \rho_G/\Theta$ with the W boson mass density $\rho_G =
M_W^4 \approx 10^8 \ \textup{GeV}^4$ at the weak scale and
baryon mass generation by $\rho_B \equiv \Omega_B \rho_c \simeq 10^{-61}
\Omega_B \rho_G/\Theta$ with the gluon mass density $\rho_G \approx 10^{-2}
\ \textup{GeV}^4$ at the strong scale. $\Theta$ terms as the
surface terms modify the original GWS model for weak interactions
and the original QCD for strong interactions, which have problems
in fermion mass terms violating gauge invariance, and suggest
mass generation as the nonperturbative breaking of gauge and
chiral invariance through DSSB.

In this approach, vacuum energy and matter energy are spatially quantized as well as
photon energy and phonon energy. The vacuum represented by massive gauge bosons is
quantized by the maximum wavevector mode $N_R = i/(\Omega -1)^{1/2} \approx 10^{30}$
and the total gauge boson number $N_G = 4 \pi N_R^3/3 \approx 10^{91}$. The maximum
wavevector mode $N_R \approx 10^{30}$ is manifest since the universe size is $R_{Pl}
\approx 10^{-3}$ cm at the Planck scale $l_{Pl} \approx 10^{-33}$ cm and the universe
size is $R_{0} \approx 10^{28}$ cm at the present scale $l_{Pl} \approx 10^{-3}$ cm if
the universe is extremely flat, $\Omega - 1 = - 10^{-61}$. Baryon matter represented
by massive baryons is quantized by the maximum wavevector mode (Fermi mode) $N_F
\approx 10^{26}$ and the total baryon number $B = N_B = 4 \pi N_F^3/3 \approx
10^{78}$. Baryon matter quantization is consistent with the nuclear matter number
density $n_n \approx n_B \approx 1.95 \times 10^{38} \ \textup{cm}^{-3}$ and
Avogadro's number $N_A = 6.02 \times 10^{23} \ \textup{mol}^{-1} \approx 10^{19} \
\textup{cm}^{-3}$ in the matter; the baryon number density at the nuclear interaction
scale $10^{-1}$ GeV is $10^{26} \ \textup{cm}^{-3}$ in the universe size $R_{QCD}
\approx 10^{17}$ cm, whose volume $10^{51} \ \textup{cm}^{3}$ is $10^{12}$ times
bigger than the matter volume $10^{39} \ \textup{cm}^{3}$. Electrons with the mass
$0.5$ MeV might be similarly quantized by $N_F \approx 10^{27}$ and the total number
$10^{81}$ if the electron number is conserved under the assumption of $\Omega_e =
\rho_e/\rho_c \approx 1$: since the baryon number minus the lepton number ($B - L$),
the baryon number, and the lepton number seem to be conserved below the weak energy,
the total electron number $10^{81}$ different with the total baryon number $10^{78}$
suggests lepton matter as dark matter. The maximum wavevector mode $N_F$ is close to
$10^{30}$ if the mass quantization unit of fermions $10^{-12}$ GeV is used rather than the
mass unit of baryons $0.94$ GeV under the assumption of the fermion number conservation.
Massless photons are quantized by the maximum
wavevector mode $N_\gamma \approx 10^{29}$ and the total photon number $N_{t \gamma} =
4 \pi N_\gamma^3/3 \approx 10^{88}$. CMBR is the crucial evidence for massless gauge
bosons (photons) with the number $N_{t \gamma} \approx 10^{88}$. Massless phonons in
the matter space are quantized by the maximum wavevector mode
(Debye mode) $N_D \approx 10^{25}$ and the total phonon number $N_{t p} = 4 \pi
N_D^3/3 \approx 10^{75}$. These total particle numbers $N_G \approx 10^{91}, N_B
\approx 10^{78}$, $N_{t \gamma} \approx 10^{88}$, and $N_{t p} \approx 10^{75}$ are
conserved good quantum numbers. Vacuum, matter, photon, and phonon energies are also
thermodynamically quantized.  Quantum states of vacuum, matter, photons, and phonons
have average occupation numbers $f_b = 1 /(e^{(E - \mu)/T} - 1)$ for gauge bosons,
$f_f = 1 /(e^{(E - \mu)/T} + 1)$ for baryons, $f_\gamma = 1 /(e^{E/T} - 1)$ for
photons, and $f_p = 1 /(e^{E/T} - 1)$ for phonons under the assumption of free
particles in thermal equilibrium.

Total particle numbers such as the gauge boson number $N_G \approx 10^{91}$, the
baryon number $N_B \approx 10^{78}$, the electron number $N_e \approx 10^{81}$, the
photon number $N_{t \gamma} \approx 10^{88}$, and the phonon number $N_{t p} \approx
10^{75}$ are conserved good quantum numbers as described above. Conservation laws of
the baryon number, lepton number, and electric charge number are good in weak, strong,
and present interactions but they would be nonperturbatively violated in gravitational
interactions: there is possibility for such violation even at much lower energy
although they are extremely small. Discrete symmetries such as parity (P), charge
conjugate (C), time reversal (T), and charge conjugate and parity (CP) are conserved
perturbatively but are violated nonperturbatively during DSSB. The violation is
analogous to the nonconservation of the (V + A) current in weak interactions and to
the axial vector current in strong interactions. If baryon and electron numbers $N_B
\simeq N_e \simeq 10^{78}$ are simultaneously conserved, the mass density $\rho_B
\simeq 10^3 \ \rho_e$ contradicts with $\rho_B \approx \rho_e \approx \rho_c$ at the
Planck scale. This implies the nonconservation of baryon and lepton numbers at the
Planck scale. The conservation of the fermion number $N_f \simeq 10^{91}$ in the unit
of mass $10^{-12}$ GeV is good at the Planck scale under the assumption with no
supersymmetry and higher dimensions; quarks and leptons are postulated as composite
particles composed  from more fundamental particles. The matter current at the Planck
scale, the (V - A) current and the electromagnetic current at the electroweak scale,
the baryon current and the proton current at the strong scale, and the lepton current
at the present scale might be conserved currents at different energy scales. The
proton number conservation is the result of the $U(1)_f$ gauge theory just as the
electron number conservation is the result of the $U(1)_e$ gauge theory. The baryon
number conservation is the consequence of the $U(1)_Z$ gauge theory for strong
interactions just as the lepton number conservation is the consequence of the $U(1)_Y$
gauge theory for weak interactions. Nonperturbative symmetry breaking described above
might suggest the restoration of perfect symmetry before DSSB if supersymmetry between
fermions and bosons, higher dimensions above one-time and three-space dimensions,
symmetry between space and time, duality between intrinsic-extrinsic space, duality
between electricity-magnetism, and duality between discrete symmetries are adopted.

Significant quantum tests of QG, which are compatible with BUMERANG-98 and MAXIMA-1,
are summarized as follows. The DSSB of local gauge symmetry and global chiral symmetry
triggers the baryon current anomaly. The relation of QG with the inflation theory is
analyzed; the vacuum energy relevant for the gauge boson mass is the source of
inflation and the universe is flat always. In addition, a gauge theory responsible for
the expansion of the present universe is suggested and the evidence of the ongoing
phase transition is CMBR. Nearly massless gauge bosons with the mass $10^{-12}$ GeV
are considered to be mediators of nonbaryonic dark matter: SIMPs are suggested as
observable dark matter in addition to WIMPs. The structure formation, baryon
asymmetry, matter mass generation, and nucleosynthesis are consistent with this
scheme. Fermion mass generation and $\Theta$ vacuum are resolved in terms of intrinsic
and extrinsic quantum numbers. The proton number conservation is the consequence of
the $U(1)_f$ gauge theory just as the electron number conservation is the consequence
of the $U(1)_e$ gauge theory and the baryon number conservation is the result of the
$U(1)_Z$ gauge theory for strong interactions just as the lepton number conservation
is the result of the $U(1)_Y$ gauge theory for weak interactions. Duality between
intrinsic and extrinsic spacetime or duality between electricity and magnetism is
suggested as one of underlying principles before DSSB. The relation between time and
gauge boson mass is introduced. The potential QG as a gauge theory may resolve serious
problems of GUTs and the standard model: different gauge groups, Higgs particles, the
inclusion of gravity, the proton lifetime, the baryon asymmetry, the family symmetry
of elementary particles, inflation, fermion mass generation, etc. This scheme may also
provide possible resolutions to the problems of Einstein's general relativity or the
standard hot big bang theory: the spacetime singularity, cosmological constant,
quantization, baryon asymmetry, structure formation, dark matter, flatness of the
universe, and renormalizability, etc.  This work toward QG would thus shed light on
understanding fundamental forces in nature and its consequences play significant roles
in various fields since all the materials in nature over all length scales are more or
less governed by fundamental forces generated by QG.

\onecolumn

\begin{table}
\caption{\label{fuco} Fundamental and Cosmological Constants in
Quantum Cosmology}
\end{table}
\centerline{
\begin{tabular}{|c|c|c|c|c|} \hline
Constant  & Gravity  & Weak & Strong & Present \\ \hline \hline
Gauge Boson Mass $M_G$ (GeV) & $10^{19}$ & $10^2$ & $10^{-1}$ & $10^{-12}$ \\ \hline
Effective Coupling Constant $G_G$ ($\textup{GeV}^{-2}$) & $10^{-38}$ &  $10^{-5}$
& $10^{-1}$ & $10^{24}$ \\ \hline
Gauge Boson Number Density $n_G$ ($\textup{cm}^{-3}$) & $10^{98}$ & $10^{47}$ & $10^{39}$ & $10^{5}$ \\ \hline
Vacuum Energy Density $V_e$ ($\textup{g} \ \textup{cm}^{-3}$) & $10^{93}$ &  $10^{25}$ & $10^{14}$ & $10^{-29}$ \\ \hline
Cosmological Constant $\Lambda_e$ ($\textup{GeV}^2$) & $10^{38}$ & $10^{-30}$ & $10^{-42}$ & $10^{-84}$ \\ \hline
Hubble Constant $H_e$ ($\textup{GeV}$) & $10^{19}$ & $10^{-15}$ & $10^{-21}$ & $10^{-42}$  \\ \hline
Baryon Number Density $n_B$ ($\textup{cm}^{-3}$) & & & $10^{26}(10^{38})$ & $10^{-6}(10^{4})$ \\ \hline
Baryon Mass Density $\rho_B$ ($\textup{g} \ \textup{cm}^{-3}$) & &  & $10^{1}(10^{14})$ & $10^{-31}(10^{-20})$ \\ \hline
Electron Number Density $n_e$ ($\textup{cm}^{-3}$) & & $10^{38}(10^{49})$ & $10^{29}(10^{41})$ & $10^{-3}(10^{7})$ \\ \hline
Electron Mass Density $\rho_e$ ($\textup{g} \ \textup{cm}^{-3}$) & & $10^{23}(10^{35})$ & $10^{1}(10^{14})$ & $10^{-31}(10^{-20})$ \\ \hline
Photon Energy $E_\gamma$ ($\textup{GeV}$) & $10^{18}$ & $10^{1}$ & $10^{-2}$ & $10^{-13}$ \\ \hline
Photon Number Density $n_{t \gamma}$ ($\textup{cm}^{-3}$) & $10^{95}$ & $10^{44}$ & $10^{36}$ & $10^{2}$ \\ \hline
Photon Energy Density $\epsilon_\gamma$ ($\textup{g} \ \textup{cm}^{-3}$) &
$10^{89}$ & $10^{21}$ & $10^{10}$ & $10^{-34}$ \\ \hline
Phonon Number Density $n_{t p}$ ($\textup{cm}^{-3}$) & & & $10^{24}$ & $10^{-10}$ \\ \hline
Phonon Energy Density $\epsilon_p$ ($\textup{g} \ \textup{cm}^{-3}$) & & & $10^{-2}$ & $10^{-46}$ \\ \hline
Constant $\Theta$ & $10^{61}$ & $10^{-4}$ & $10^{-12}$ & $10^{-61}$ \\ \hline
Topological Constant $\nu$ & $10^{-122}$ & $10^{-57}$ & $10^{-49}$ & $10^{0}$ \\ \hline
\end{tabular}
}


\begin{table}
\caption{\label{cola} Overview of Conservation Laws}
\end{table}
\centerline{
\begin{tabular}{|c|c|c|c|c|} \hline
Conservation  & Gravity & Electromagnetic & Weak & Strong \\ \hline \hline Energy,
Momentum, Angular Momentum & yes & yes & yes & yes \\ \hline Charge, Baryon, Lepton &
no & yes & yes & yes \\ \hline P, C, T, CP & no & yes & no & no \\ \hline TCP & yes &
yes & yes & yes \\ \hline
\end{tabular}
}


\begin{table}
\caption{\label{coga} Relations between Conservation Laws and
Gauge Theories}
\end{table}
\centerline{
\begin{tabular}{|c|c|c|} \hline
Force & Conservation Law & Gauge Theory \\ \hline \hline
Electromagnetic & Proton & $U(1)_f$ \\ \hline
Strong & Baryon & $U(1)_Z$ \\ \hline
Strong & Color Vector & $SU(2)_N \times U(1)_Z$ \\ \hline
Strong & Color & $SU(3)_C$ \\ \hline
Electromagnetic & Electron & $U(1)_e$ \\ \hline
Weak & Lepton & $U(1)_Y$ \\ \hline
Weak & V-A & $SU(2)_L \times U(1)_Y$ \\ \hline
Weak & Isotope (isospin) & $SU(3)_I$ \\ \hline
\end{tabular}
}


\begin{table}
\caption{\label{scqc} Scales in Quantum Cosmology}
\end{table}
\centerline{
\begin{tabular}{|c|c|c|c|c|} \hline
Scale & Time t (s) & Energy E (GeV) & Temperature T (K) & Universe Size R (cm) \\ \hline \hline
Planck $l_{Pl}$ & $10^{-43}$ & $10^{19}$ & $10^{32}$ & $10^{-3}$ \\ \hline
Weak $l_{EW}$ & $10^{-10}$ & $10^{2}$ & $10^{15}$ & $10^{14}$ \\ \hline
Strong $l_{QCD}$ & $10^{-5}$ & $10^{-1}$ & $10^{12}$ & $10^{17}$ \\ \hline
Today $l_0$ & $10^{17}$ & $10^{-12}$ & $3$ & $10^{28}$ \\ \hline
\end{tabular}
}

\end{document}